\documentclass[pdflatex,sn-mathphys-num]{sn-jnl}

\usepackage{graphicx}%
\usepackage{multirow}%
\usepackage{amsmath,amssymb,amsfonts}%
\usepackage{amsthm}%
\usepackage{mathrsfs}%
\usepackage[title]{appendix}%
\usepackage{xcolor}%
\usepackage{textcomp}%
\usepackage{manyfoot}%
\usepackage{booktabs}%
\usepackage{algorithm}%
\usepackage{algorithmicx}%
\usepackage{algpseudocode}%
\usepackage{listings}%
\usepackage{makecell}
\usepackage{subcaption}
\usepackage{braket}

\theoremstyle{thmstyleone}%
\theoremstyle{thmstyletwo}%

\theoremstyle{thmstylethree}%

\begin{document}

\title[Article Title]{Large Scale Entanglement Structure Detection in 100-Qubit Systems via Local Joint Measurements }


\author[1]{\fnm{Rui} \sur{Li}}

\author[2]{\fnm{Yuhang} \sur{Wang}}

\author[3]{\fnm{Chunxiao} \sur{Du}}

\author[4]{\fnm{Shikun} \sur{Zhang}}

\author[5,6]{\fnm{Zheng} \sur{Qin}}

\author[7]{\fnm{Wenxiu} \sur{Li}}

\author[8]{\fnm{Hao} \sur{Zhang}}
\author*[2,3]{\fnm{Zhisong} \sur{Xiao}}\email{zsxiao@buaa.edu.cn}

\affil*[1]{\orgdiv{School of Applied Science}, \orgname{Beijing Information Science and Technology University}, \orgaddress{\city{Beijing 100192}, \country{China}}}

\affil[2]{\orgdiv{School of Instrument Science and Opto-Electronics Engineering}, \orgname{Beijing Information Science and Technology University}, \orgaddress{\city{Beijing 100192}, \country{China}}}

\affil[3]{\orgdiv{School of Physics}, \orgname{Beihang University}, \orgaddress{\city{Beijing 100191}, \country{China}}}

\affil[4]{\orgdiv{School of Future Technology}, \orgname{Henan University}, \orgaddress{\city{Zhengzhou 450046}, \country{China}}}

\affil[5]{\orgdiv{}\orgname{Shenzhen Institute of Beihang University}, \orgaddress{\city{Shenzhen 518063}, \country{China}}}

\affil[6]{\orgdiv{} \orgname{Jiangxi Beidouyun Intelligent Technology Co. Ltd.}, \orgaddress{\city{Nanchang 330038}, \country{China}}}

\affil[7]{\orgdiv{School of Automation (School of Artificial Intelligence)}, \orgname{Beijing Information Science and Technology University}, \orgaddress{\city{Beijing 100192}, \country{China}}}

\affil[8]{\orgdiv{School of Space and Earth Sciences}, \orgname{Beihang University}, \orgaddress{\city{Beijing 100191}, \country{China}}}


\abstract{
Identifying the entanglement structure of a many-body quantum state, namely how its constituents partition into unentangled blocks, is a central task in quantum information science, yet conventional tomography scales exponentially with system size. Here we introduce a scalable framework that recognizes large-scale entanglement structures directly from local correlation fingerprints. By choosing a representative local Pauli basis that satisfies a boundary-matching condition $p_1 = p_R$, the entire chain is read out in a single measurement configuration, keeping the measurement effort independent of system size. In noisy simulations, this single-basis protocol classifies GHZ-, W-, and cluster-type structures among 30 candidate partitions with a mean accuracy exceeding 95\% for systems of up to 100 qubits. 
We further validate the protocol on a superconducting quantum processor, where it reliably classifies block structures for systems of up to 13 qubits before noise- and depth-induced degradation sets in at larger sizes. By mapping these failure modes explicitly, our results delineate the boundary of hardware-level scalability and point to a concrete strategy for characterizing entanglement structure on near-term quantum devices.}




\maketitle

\section{Introduction}


Multipartite entanglement is a defining resource of quantum information science \cite{bib1, bib2}, underlying advantages in quantum computation and information processing \cite{bib3, bib6}, communication and metrology \cite{bib7}, and serving as a central diagnostic for intermediate-scale quantum devices \cite{bib4,bib5}. For large processors, however, merely certifying the presence of entanglement is rarely sufficient. Many advanced applications depend heavily on the finer entanglement structure, namely how qubits partition into mutually separable blocks \cite{bib8, bib12}, how deep the entanglement extends \cite{bib11}, and which groups of particles form correlated units \cite{bib9,bib10}. The conventional pathway to characterizing such partitions typically relies on full quantum state tomography (QST) to reconstruct the complete density matrix, whose measurement and post-processing costs scale exponentially with the system size $N$~\cite{bib13,bib14,bib15}. This scaling makes the direct characterization of large-scale entanglement structures a major challenge.
 
To reduce the measurement cost of traditional QST, several advanced tomographic paradigms, including compressed sensing~\cite{bib16}, tensor-network~\cite{bib17,bib18,bib19}, and neural-network tomography~\cite{bib20,bib21,bib22,bib23,bib24}, have been developed to mitigate this burden for structured states, yet they still aim to represent the full state~\cite{bib26,bib27,bib28,bib29}. On the other hand, entanglement witnesses offer a more economical alternative but typically require prior knowledge of the target state and must be redesigned for different structures or state families~\cite{bib4,bib5,bib8,bib12}. These limitations motivate the development of alternative methods that can infer entanglement structures directly from experimentally accessible data without state reconstruction.

Randomized measurements and data-driven inference have recently opened new pathways for directly probing quantum properties~\cite{bib30,bib31,bib32}. In particular, classical shadows and related protocols predict multiple linear expectation values of a state from a small, size-independent set of measurements~\cite{bib33,bib34,bib35,bib36}. Machine-learning methods have similarly been applied to state verification, phase classification, and entanglement detection~\cite{bib25,bib37,bib38,bib39,bib40}. However, a crucial distinction must be made regarding the nature of the task. Classical shadows are designed to estimate expectation values of linear observables. Detecting entanglement structures, by contrast, is a nonlinear classification problem that determines membership over structured convex sets of separable partitions. Consequently, the favorable sampling guarantees of shadow-based estimation do not directly transfer, requiring dedicated strategies. Recently, a multiview neural-network approach was introduced to resolve entanglement partitions using a limited number of global Pauli measurements~\cite{bib41,bib42}. However, that method records global $N$-qubit outcome distributions of dimension $2^N$, where the number of informative global settings scales rapidly with system size, limiting practical demonstrations to small systems ($N \le 19$). This bottleneck motivates shifting entanglement characterization from global measurements to the local level, thereby avoiding the exponential scaling of the full Hilbert space.

This shift is physically motivated by the observation that block partitions leave characteristic signatures in local correlations. For GHZ-type blocks, the two-point correlation $\langle Z_i Z_j \rangle$ is close to unity inside a block and drops sharply across its boundary, imprinting the partition directly onto short-range $Z$-type correlations~\cite{bib43,bib44}. For W-type blocks, the delocalized single excitation produces distinct local occupation and transverse correlation patterns~\cite{bib45,bib46}, whereas stabilizer-like correlations play the same role for cluster-type blocks~\cite{bib47,bib48,bib49,bib50}. This locality principle aligns with recent evidence that global properties of structured many-body states can be learned efficiently from short-range correlations~\cite{bib51}.Although some structural features are genuinely nonlocal, we frame the task precisely as the supervised recognition of the structure label within a physically motivated, locally distinguishable ensemble. Under this guiding principle, we introduce a scalable framework for large-scale entanglement structure detection based on local correlation signatures. Our protocol partitions an $N$-qubit chain into overlapping local subsystems of $R$ contiguous qubits~\cite{bib28}. By applying a translationally uniform local Pauli basis across all subsystems, the outcome space of each subsystem is capped at $2^R$ rather than $2^N$, and the total number of distinct measurement configurations remains independent of system size~\cite{bib30,bib35}, while the classifier input feature dimension grows only linearly, $\mathcal{O}(N)$.

Our analysis reveals that the structural information is redundantly distributed across local Pauli bases rather than concentrated in a few finely-tuned configurations. We show that a single well-chosen local basis ($K_{\mathrm{local}} = 1$) already suffices for highly accurate structure recognition, which represents a decisive advantage for practical experimental deployment~\cite{bib42}. To emulate realistic experimental conditions, all training and testing datasets in the main experiments are generated with rotation-gate noise; depolarizing and white noise are applied only in the dedicated stress test~\cite{bib37}. We demonstrate that our method successfully classifies GHZ-, W-, and cluster-type block structures among 30 candidate partitions with mean accuracy exceeding $95\%$ for systems up to $N = 100$ qubits in noisy simulation, using a single, size-independent measurement configuration. Furthermore, we experimentally validate the feasibility and robustness of this protocol on a superconducting quantum processor~\cite{bib52}, where it reliably classifies GHZ-type structures for systems up to 13 qubits; by testing sizes up to 16 qubits we systematically diagnose the noise- and depth-induced failure modes, thereby delineating the operational boundaries of hardware-level scalability. Together, these results establish local correlation signatures as a robust, experimentally viable, and scalable paradigm for entanglement characterization on near-term quantum devices.

\section{Results}\label{sec2}

\subsection{Numerical Experiments}

\begin{figure}[t]
  \centering
  \includegraphics[width=\textwidth]{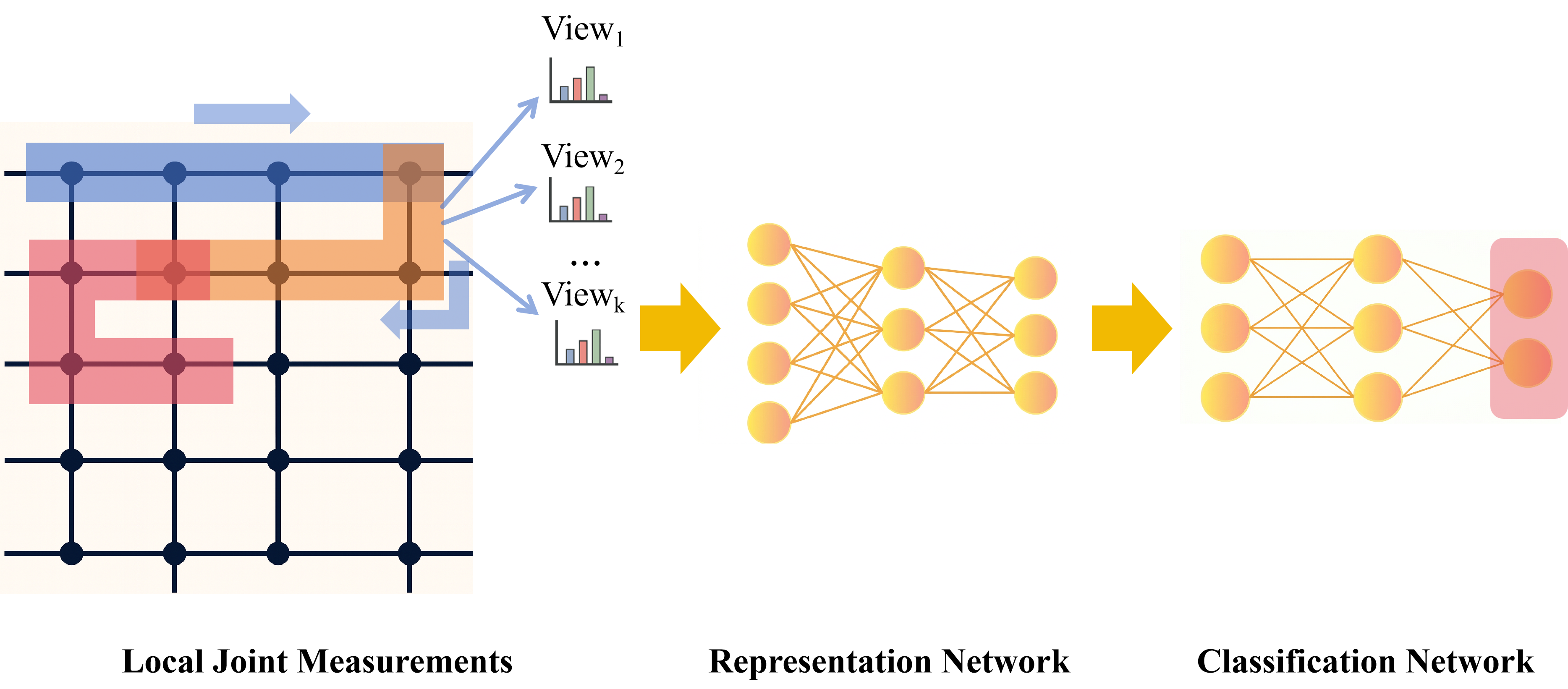}
  \caption{\textbf{Local-joint-measurement pipeline.} An $N$-qubit chain is covered by overlapping local subsystems of $R$ qubits with step size $s$; the final window is anchored to the right boundary to ensure full coverage. In each window a single local Pauli basis is measured, yielding an outcome distribution of fixed dimension $2^R$. The per-window distributions are concatenated into the feature vector $\mathbf{x}$ of dimension $N_{\mathrm{win}}2^R$, which is passed to a representation network and then a
  classification network that outputs the entanglement-structure label.}
  \label{fig:pipeline}
\end{figure}

We organize the results around a single physical claim: for structured multipartite states, the block partition is redundantly encoded in local correlations, so that the structure label can be recovered from a small, size-independent set of local joint measurements (Fig.~\ref{fig:pipeline}). 

To operationalize the local joint-measurement representation, the $N$-qubit chain is mapped onto a sequence of overlapping local subsystems, each spanning $R$ contiguous qubits with a translational step size $s$; the final window is anchored to the right boundary so that every qubit is covered (Methods). Within this scheme the total measurement and post-processing cost is governed by three decoupled quantities: (i) the number of observation windows $N_{\mathrm{win}}$, which grows only linearly with $N$; (ii) the number of local Pauli bases measured per window, $K_{\mathrm{local}}$, which sets the measurement effort per window; and (iii) the dimension of each per-window outcome distribution, fixed at $2^R$ irrespective of $N$. This invariant $2^R$ support is the physical mechanism that circumvents the exponential sampling cost of global schemes: the samples needed to resolve each local distribution scale with the constant window size $R$ rather than with the total system size. Combined with the linear growth of $N_{\mathrm{win}}$, the total measurement cost therefore scales only as $\mathcal{O}(N)$, establishing a scalable route to entanglement characterization on large near-term processors.

\begin{figure}[t]
  \centering
  \includegraphics[width=\textwidth]{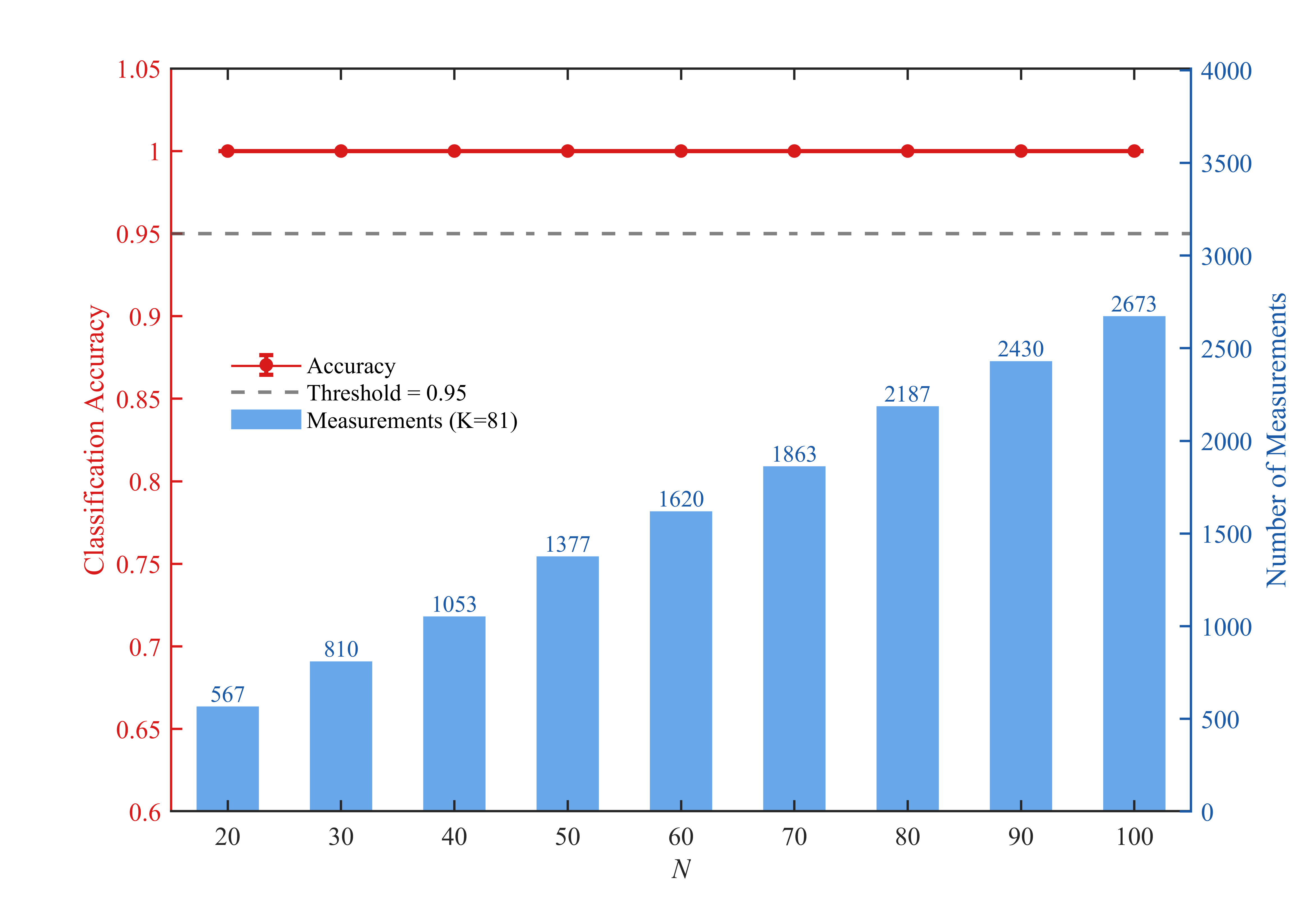}
  \caption{\textbf{Classification accuracy and measurement cost with the full local Pauli basis set.} Test accuracy (red, left axis) and total number of measurements (blue bars, right axis) for $N=20,30,\dots,100$ at fixed $R=4$, $s=3$, with all $K=81$ local bases retained. Accuracy is exactly 100\% across all $N$, well above the 0.95 threshold (dashed grey line), while the measurement count grows monotonically from 567 to 2673.}
  \label{fig:fullbasis}
\end{figure}

Before turning to the question of how many local bases are actually necessary, we first establish a reference point using the complete local measurement setting. As shown in Fig.~\ref{fig:fullbasis}, we train and test the classifier using all $K=81$ local Pauli bases. In this case, the classifier achieves perfect classification accuracy over the whole range $N=20,\dots,100$, showing that when the complete set of local Pauli measurements is available, the local outcome distributions contain sufficient information to distinguish the underlying entanglement structures. However, this performance comes at a rapidly increasing measurement cost. At fixed window size $R=4$ and shift $s=3$, the total number of measurement instances grows from 567 at $N=20$ to 2673 at $N=100$, because each local basis must be measured on every local window.

The perfect accuracy thus prompts a sharper representational question: does the network merely exploit an overcomplete, high-dimensional feature space, or are the block boundaries faithfully imprinted on a much smaller subset of local correlations? Resolving this dichotomy is essential for practical hardware deployment, since it determines whether the measurement budget can be substantially reduced without sacrificing accuracy. We therefore next examine how the performance changes when only a subset of local Pauli bases is retained.

\begin{figure}[t]
  \centering
  \includegraphics[width=\textwidth]{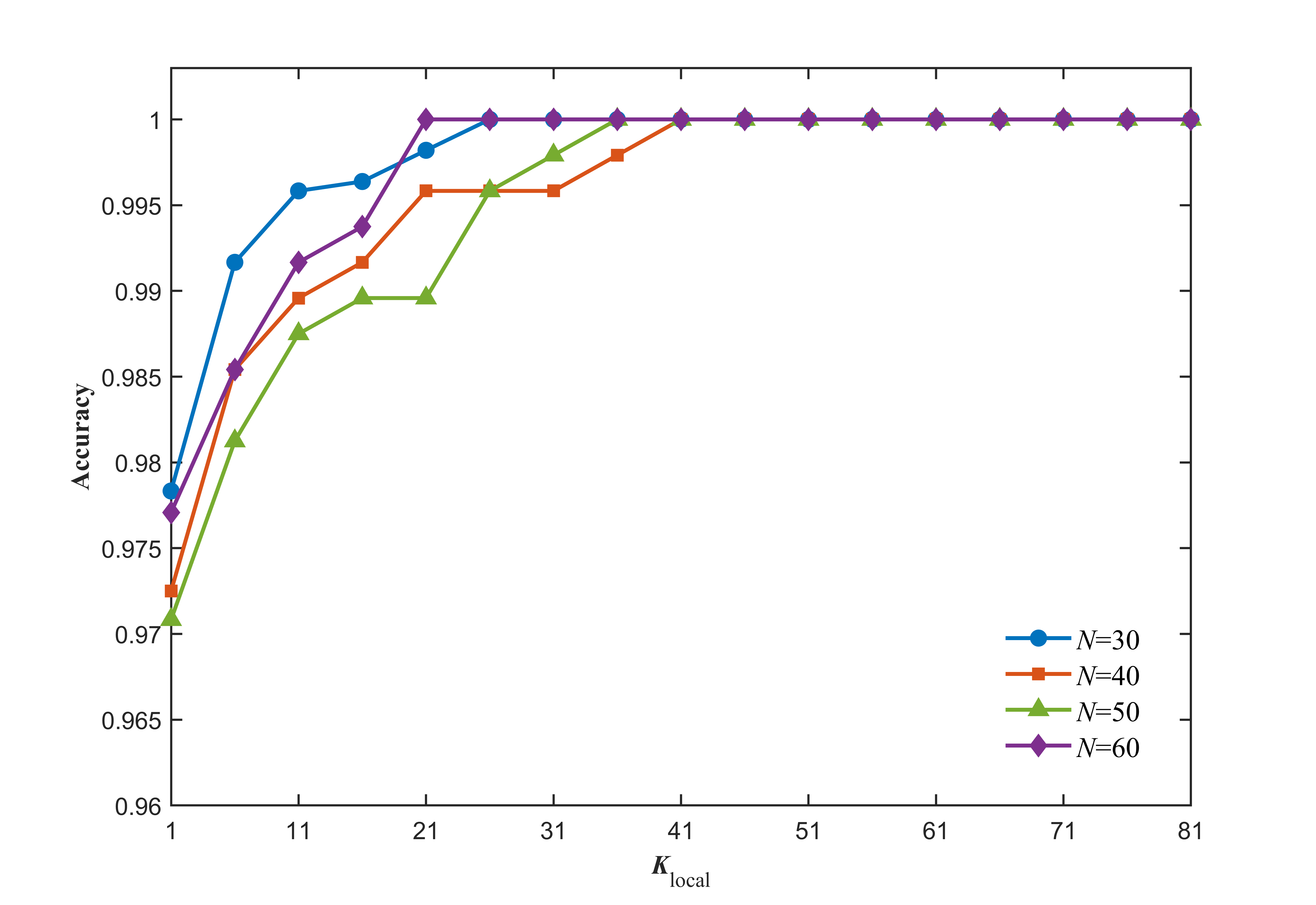}
  \caption{\textbf{Classification accuracy versus the number of retained local Pauli bases $K_{\mathrm{local}}$.} Test accuracy for $N=30,40,50,60$ at fixed $R=4$, $s=3$. Accuracy already exceeds 0.97 at $K_{\mathrm{local}}=1$ and saturates by $K_{\mathrm{local}}\approx 40$.}
  \label{fig:klocal}
\end{figure}

To resolve this, we systematically prune the feature space by restricting the number of retained local Pauli bases $K_{\mathrm{local}}$ per window. As shown in Fig.~\ref{fig:klocal}, the accuracy rises monotonically with $K_{\mathrm{local}}$ and saturates at $100\%$ near $K_{\mathrm{local}}\simeq 40$, yet it is already high in the single-basis limit: at $N=60$ a single properly chosen basis exceeds $97\%$. This rapid saturation shows that the signatures of the block boundaries are redundantly encoded across a broad set of local Pauli bases, so that the full 81-basis representation is largely overcomplete. Rather than depending on a finely tuned configuration, the spatial partition is robustly imprinted on the marginals of almost any representative basis. This redundancy is the cornerstone of our measurement-reduction strategy: moving from the full 81-basis set to a single basis ($K_{\mathrm{local}}=1$) compresses the number of measurement settings from $81\,N_{\mathrm{win}}$ to $N_{\mathrm{win}}$—a factor-of-81 reduction—without sacrificing classification reliability. This reduction shifts the central question from how many bases are needed to how the windows should be arranged and whether a single basis can be chosen robustly across state families.

\begin{figure}[h]
    \centering
    \begin{subfigure}{0.8\linewidth}
        \centering
        \includegraphics[width=\linewidth]{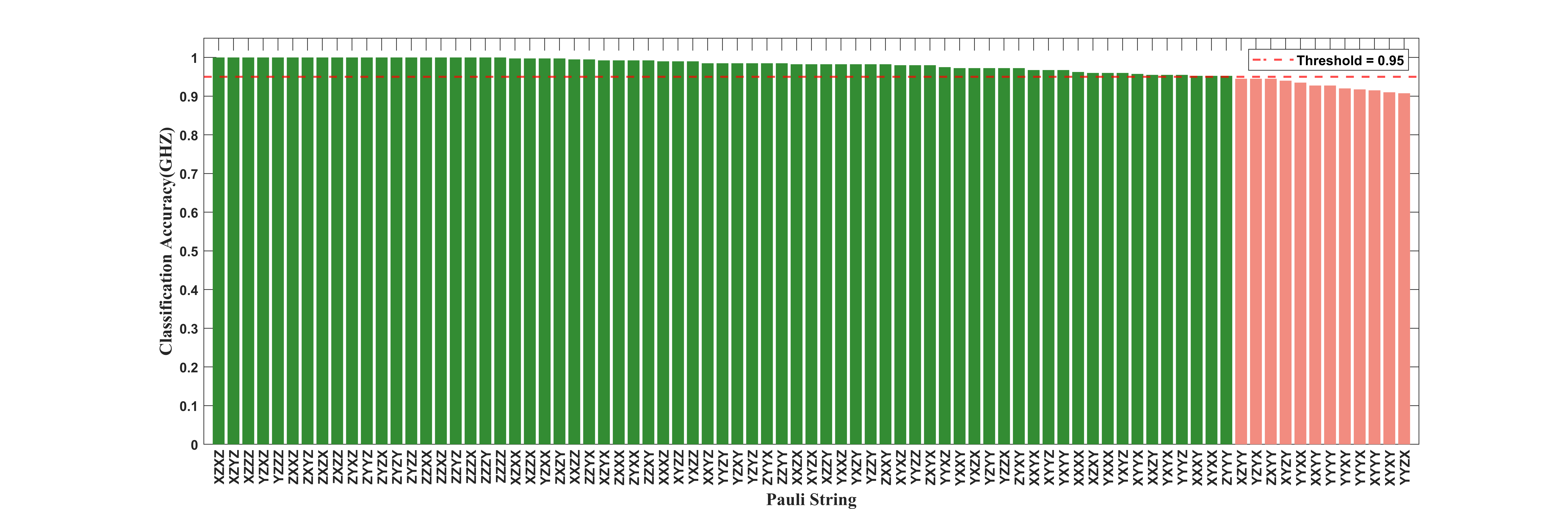}
        \caption{GHZ state}
        \label{fig:result_Kvalue_ghz}
    \end{subfigure}
    \vspace{2mm}
    \begin{subfigure}{0.8\linewidth}
        \centering
        \includegraphics[width=\linewidth]{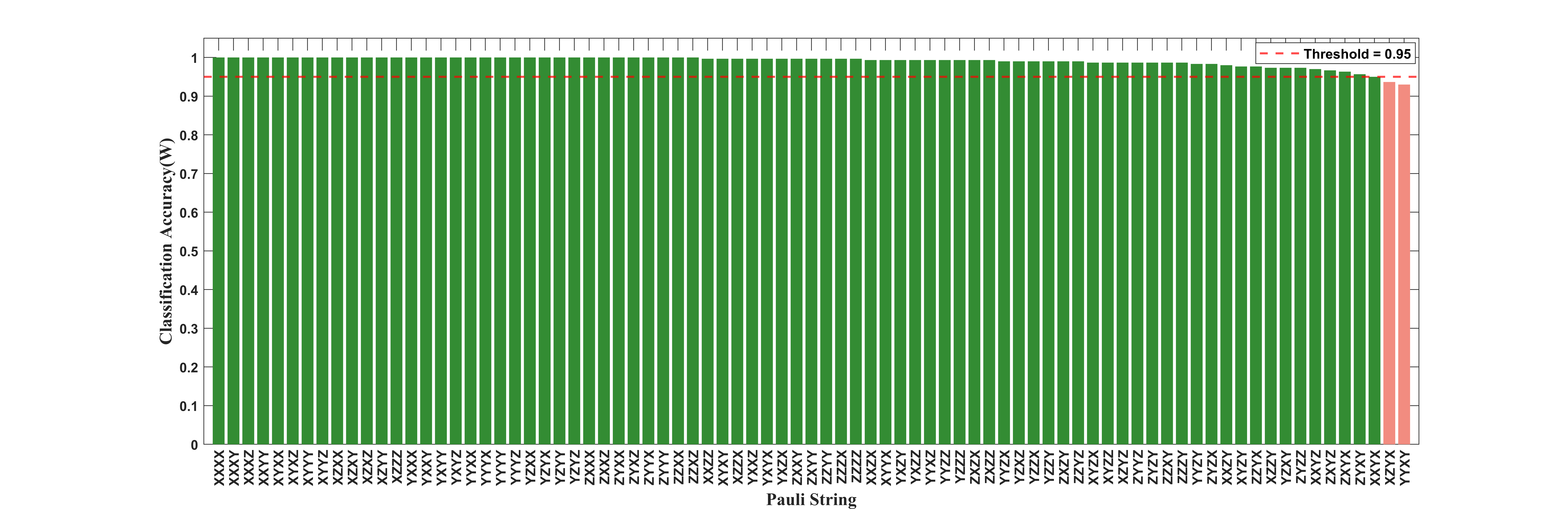}
        \caption{W state}
        \label{fig:result_Kvalue_w}
    \end{subfigure}
    \vspace{2mm}
    \begin{subfigure}{0.8\linewidth}
        \centering
        \includegraphics[width=\linewidth]{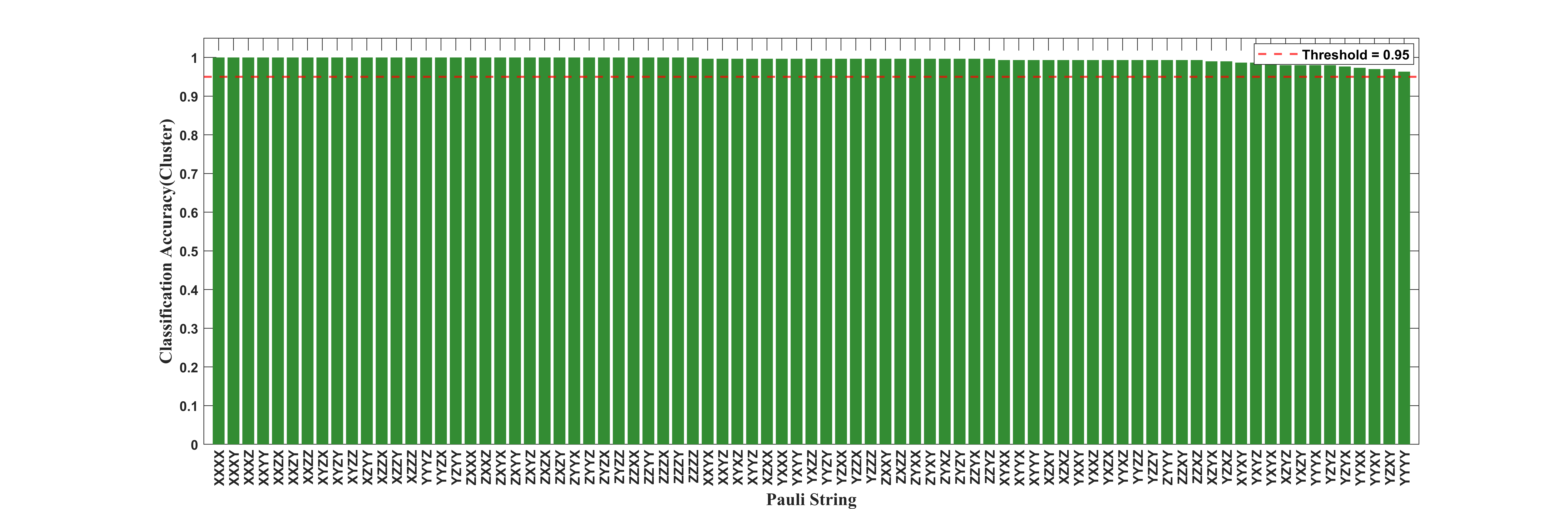}
        \caption{Cluster state}
        \label{fig:result_Kvalue_cluster}
    \end{subfigure}
    \caption{\textbf{Per-basis classification accuracy for the $3^{R}=81$ individual local Pauli bases}
    (GHZ-, W-, Cluster- block task, $N=100$, $R=4$, $s=3$). Each bar is the accuracy obtained when only that single basis is used; the dashed line marks the $0.95$ threshold, with bases above (green) and below (red) it.}
    \label{fig:perbasis}
\end{figure}

Because a single local basis is used in the reduced protocol, it is important to understand how much information is carried by different Pauli bases. We therefore examine the classification performance of all 81 local Pauli strings individually at $N=100$ and $K_{\mathrm{local}}=1$. The results for GHZ-block, W-block, and cluster-block structures are shown in Figs.~\ref{fig:perbasis}a--\ref{fig:perbasis}c. For each state family, most local Pauli bases yield accuracies above the threshold 0.95. Only a small fraction of bases fail to meet this criterion. This demonstrates that the local-correlation representation is not tied to a unique or fine-tuned measurement setting. Instead, the relevant structural information is distributed over a broad set of local Pauli measurements.

The behavior of different bases also reflects the physical correlations of the
corresponding state families. For GHZ-type structures, informative bases are associated with correlations that distinguish coherent blocks and identify their boundaries. For W-type and cluster-type structures, the useful information is encoded in different local correlation channels. Nevertheless, the same qualitative conclusion holds for all three cases: most local Pauli strings contain enough information to support reliable classification. This robustness makes the method experimentally practical, because it does not rely on an exceptionally specific measurement basis.

Since most bases are informative, a natural question is which single basis to
adopt for the subsequent experiments. To answer this, we carry out a stress test in which two additional physical noise channels are switched on. A per-qubit depolarizing channel is applied during state preparation,
\begin{equation}
\mathcal{E}_d(\rho) = (1-p_d)\,\rho
      + \frac{p_d}{3}\left(X\rho X + Y\rho Y + Z\rho Z\right),
\label{eq:depol_main}
\end{equation}
and a global white-noise channel mixes the full state with the maximally mixed
state,
\begin{equation}
\rho \longrightarrow (1-p_w)\,\rho + p_w\,\frac{I}{2^{\,N}}.
\label{eq:white_main}
\end{equation}
The depolarizing strength is swept over $p_d \in [0, 0.2]$ and the white-noise
strength over $p_w \in [0, 0.4]$. Both channels are defined at the level of the
quantum state, yet each acts exactly on any local $R$-qubit window. The depolarizing channel is unital, so it leaves the marginal of every traced-out
qubit invariant and reduces, within the window, to an independent single-qubit
stochastic map along each axis; the global white-noise channel acts as a convex
mixture with the uniform distribution,
\begin{equation}
P(s) \longrightarrow (1-p_w)\,P(s) + p_w\,\frac{1}{2^{\,R}},
\label{eq:white_marginal_main}
\end{equation}
so that both channels act exactly on every local window (see Methods).

\begin{figure}[h]
    \centering
    \begin{subfigure}{0.3\linewidth}
        \centering
        \includegraphics[width=\linewidth]{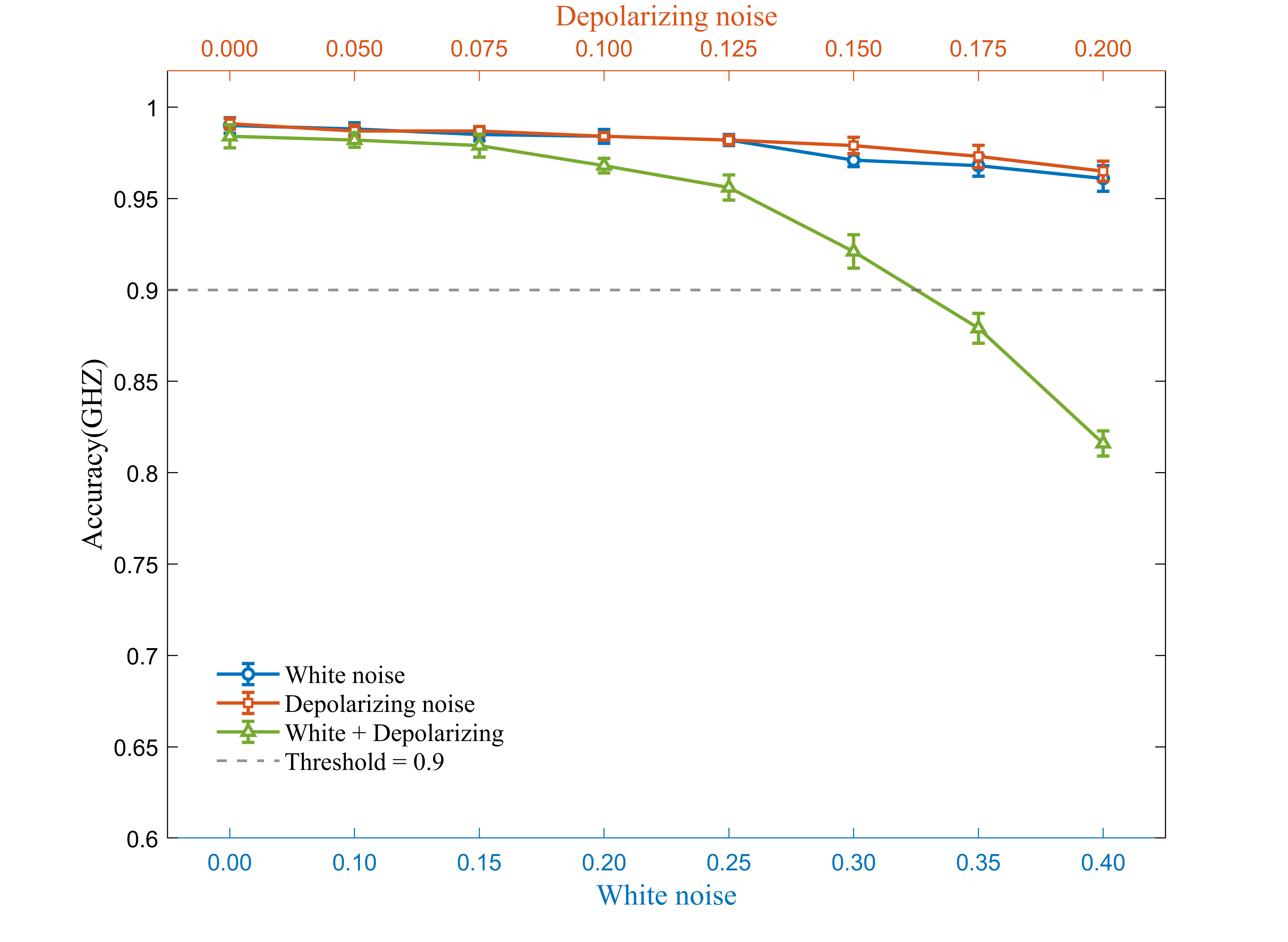}
        \caption{GHZ state}
        \label{fig:noise_ghz}
    \end{subfigure}
    \vspace{2mm}
    \begin{subfigure}{0.3\linewidth}
        \centering
        \includegraphics[width=\linewidth]{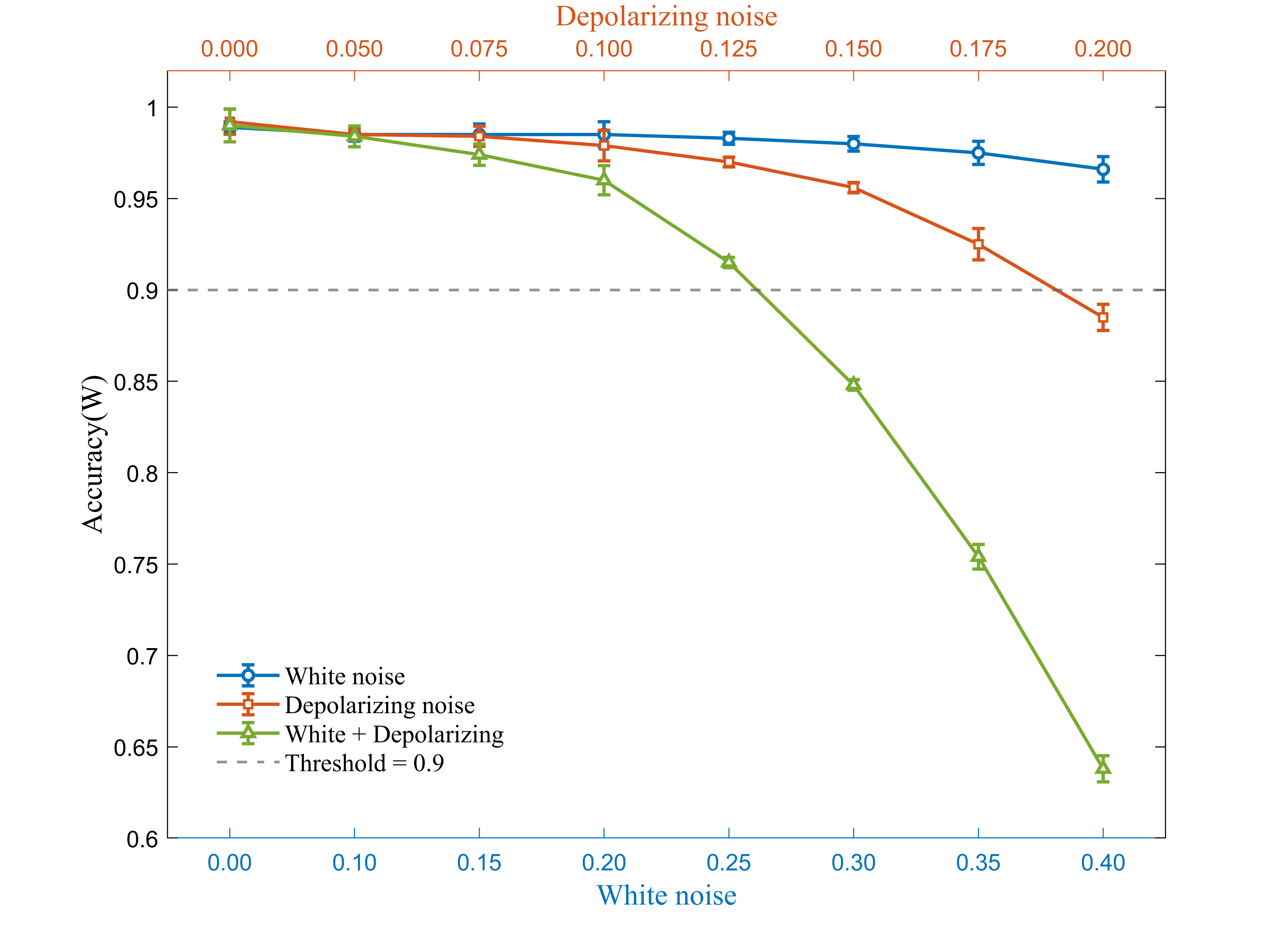}
        \caption{W state}
        \label{fig:noise_w}
    \end{subfigure}
    \vspace{2mm}
    \begin{subfigure}{0.3\linewidth}
        \centering
        \includegraphics[width=\linewidth]{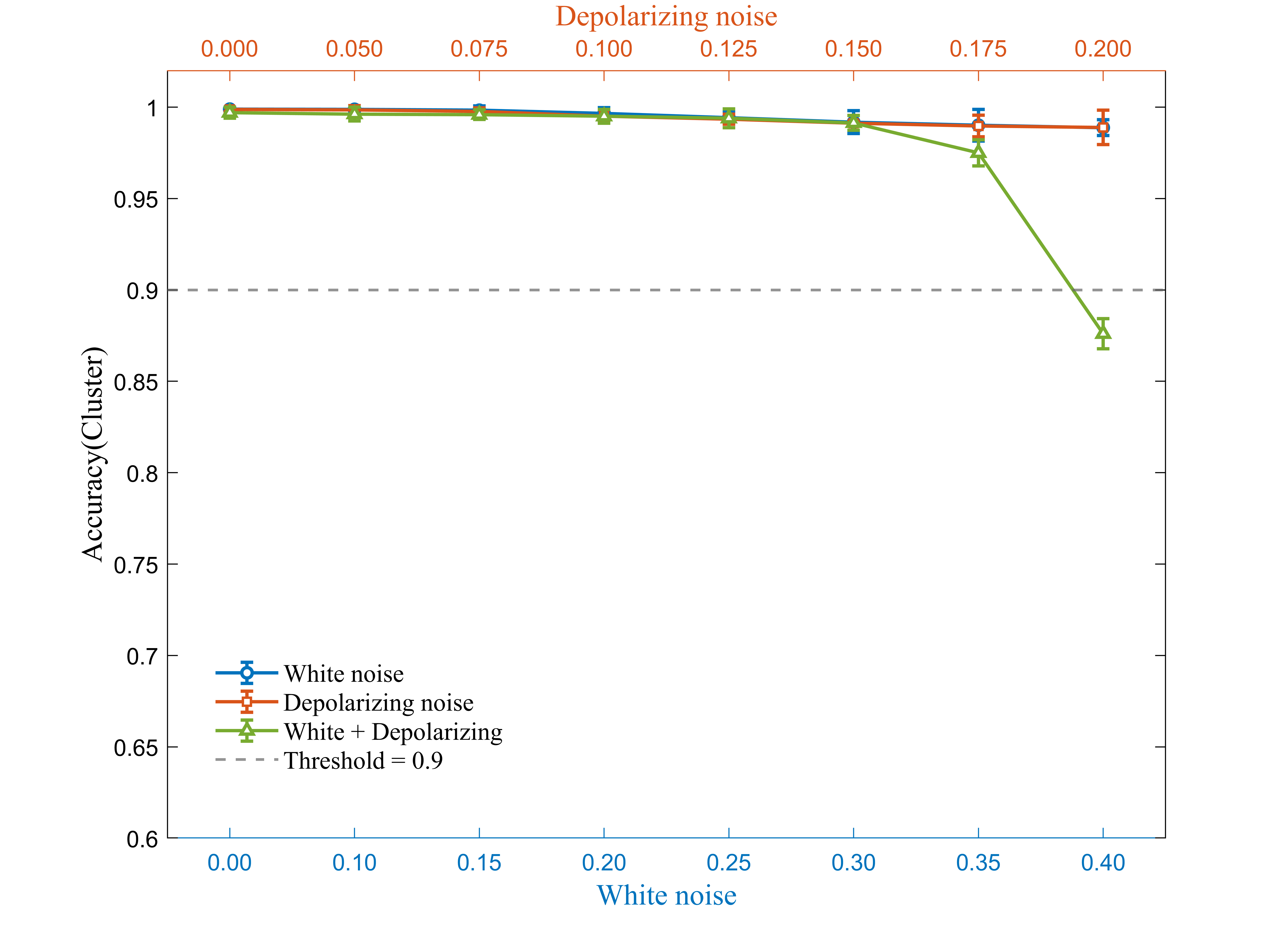}
        \caption{Cluster state}
        \label{fig:noise_cluster}
    \end{subfigure}
    \caption{\textbf{Stress test of the single-basis protocol under strong depolarizing and white noise} ($N$ fixed, $K_{\mathrm{local}}=1$). Each panel shows the classification accuracy for one state family as a function of white-noise strength $p_w$ (bottom axis, $0$--$0.4$) and depolarizing strength $p_d$ (top axis, $0$--$0.2$). Blue curves apply white noise, orange curves apply depolarizing noise, and green curves apply both simultaneously; the grey dashed line marks the 0.9 threshold and error bars denote 1 s.d.}
    \label{fig:stress}
\end{figure}

Fig.~\ref{fig:stress} reports the classification accuracy for the three state families as the depolarizing channel, the white-noise channel, and their combination are switched on. For all three families, single-channel noise is tolerated very well: with only white noise (blue) or only depolarizing noise (orange), the accuracy stays close to unity across essentially the entire swept range and remains above the 0.9 threshold even at the largest single-channel strengths. The combined channel (green) is the most demanding. The combined channel is tolerated best by cluster states (dipping only to \( \approx 0.88 \) at the most extreme setting), followed by GHZ states (\( \approx 0.82 \)), while W states degrade most steeply, falling to \( \approx 0.64 \) once both channels approach their maximal values. Two conclusions follow. First, within and moderately beyond the nominal operating regime the protocol is stable for all three families, confirming its robustness to realistic noise. Second, the sweep provides a principled way to fix the representative measurement basis: for each state family we select the basis that retains the highest accuracy across the full noise sweep while satisfying the boundary-matching condition $p_1 = p_R$ (see Methods), which yields XZYX for GHZ, YXYX for W, and ZZZX for cluster states. This matching condition is not merely a labeling convention: because adjacent windows overlap on their shared boundary qubit, requiring the first and last Pauli operators to coincide guarantees that neighboring windows agree on that qubit, so the entire $N$-qubit chain can be read out in a single hardware measurement configuration. All subsequent GHZ-type experiments therefore use the XZYX basis, and the corresponding representative bases are used for W and cluster states.

\begin{figure}[t]
  \centering
  \includegraphics[width=\textwidth]{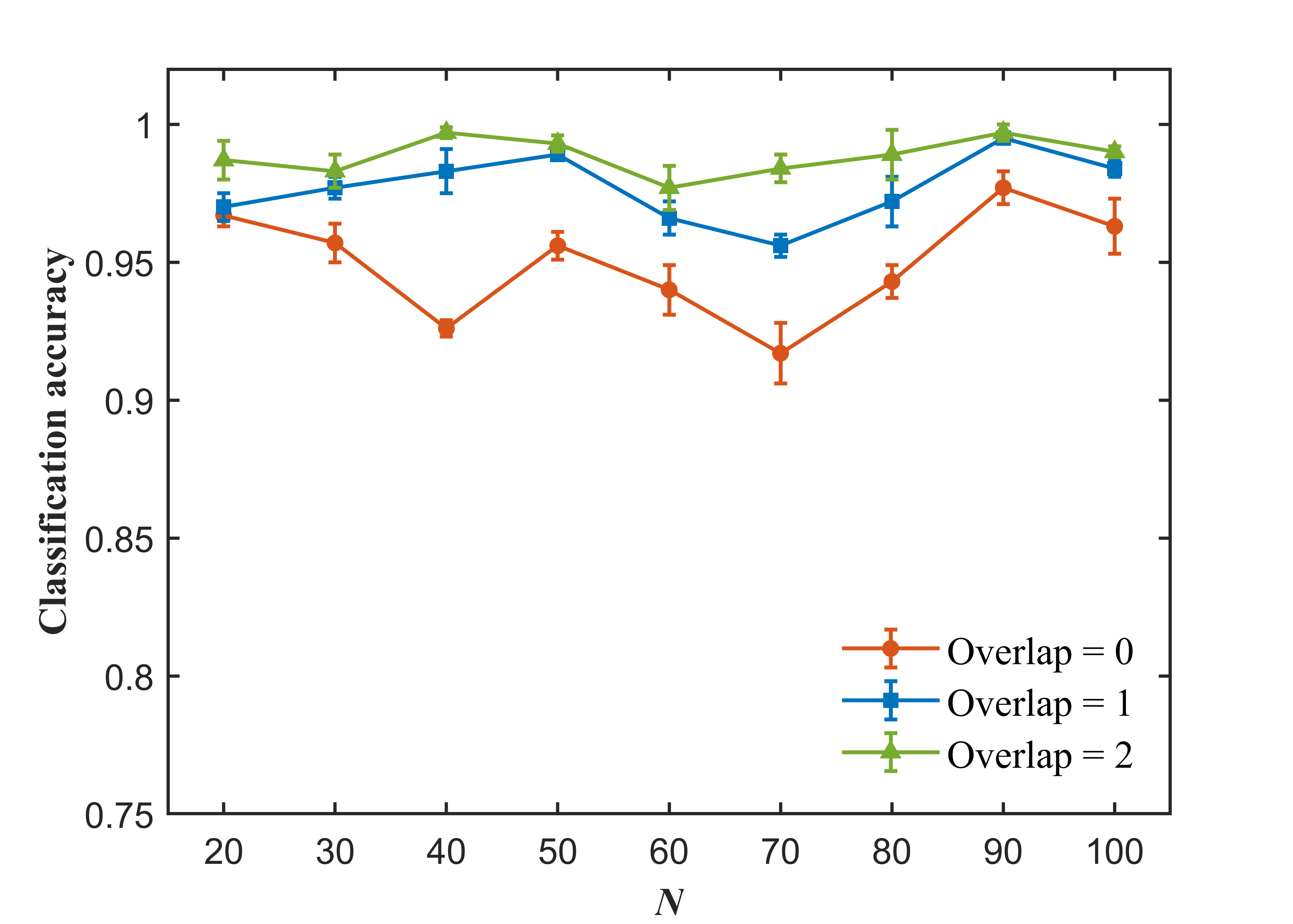}
  \caption{\textbf{Effect of window overlap on classification accuracy} (GHZ-block task, $N=20$--$100$, $R=4$, $K_{\mathrm{local}}=1$, measured in the XZYX Pauli basis). The overlap between adjacent windows equals $R-s$; overlaps of 0, 1 and 2 correspond to $s=4,3,2$. Each point is the mean over 5 independent runs and error bars denote 1 s.d. Non-overlapping local subsystems ($R-s=0$) give the lowest accuracy, and accuracy increases monotonically with overlap.}
  \label{fig:overlap}
\end{figure}

Having fixed the representative basis, we next quantify the second design parameter, the window overlap $R-s$. Non-overlapping windows, corresponding to $s=R$, tile the chain without redundancy, but a block boundary located between two neighboring windows may not be captured by any local window. As a result, correlations across such a boundary can be lost. Introducing overlap ensures that each bond is included in at least one local window.
Fig.~\ref{fig:overlap} shows the effect of different overlap values on the classification accuracy. The non-overlapping case gives the lowest accuracy, while the accuracy improves when overlap is introduced. The performance increases from overlap 0 to overlap 2, but this improvement also increases the number of local windows and hence the measurement cost.
Throughout the following analysis, we therefore adopt $R=4$ and $s=3$, corresponding to overlap 1, which provides a favorable balance between accuracy and measurement efficiency.

\begin{figure}[t]
  \centering
  \includegraphics[width=\textwidth]{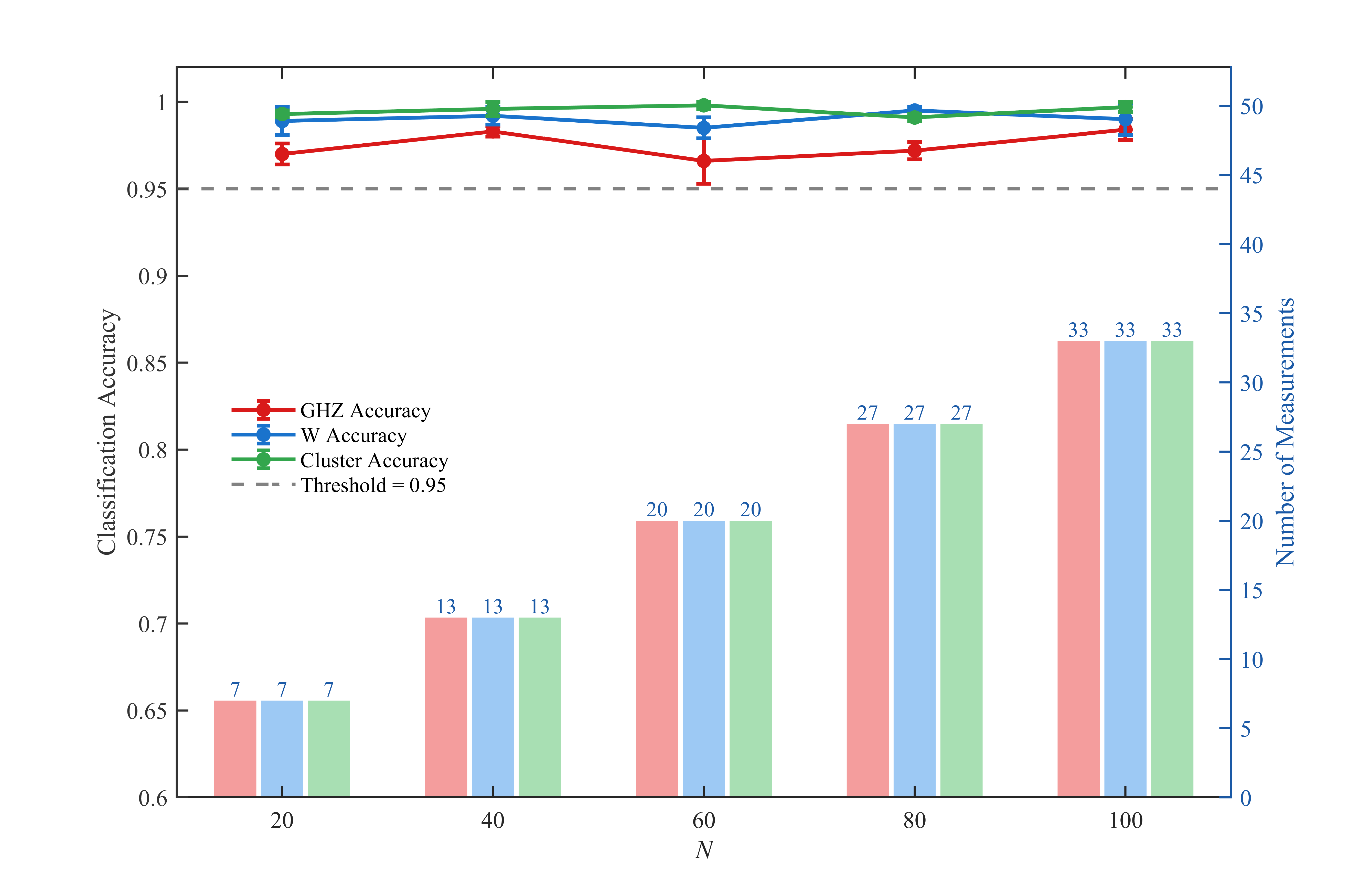}
 \caption{\textbf{Classification accuracy of the reduced protocol versus system size.}
  (GHZ, W, and cluster states, $N=20$--$100$, $K_{\mathrm{local}}=1$; each family measured in its representative basis: XZYX for GHZ, YXYX for W, ZZZX for cluster.)  Curves (left axis) show the mean accuracy for each state family, with error bars denoting 1 s.d.; the grey dashed line marks the 0.95 threshold. Bars (right axis) indicate the number of local windows $N_{\mathrm{win}}$ (7, 13, 20, 27, 33 for $N=20,40,60,80,100$). Because $K_{\mathrm{local}}=1$ and the representative basis satisfies $p_1=p_R$, all windows are acquired in a single measurement configuration; the reported bars therefore count windows, not distinct hardware circuits. Accuracy stays near or above threshold for all three families, while the single-basis setting reduces the measurement count by a factor of 81 compared with the full-basis protocol.}
  \label{fig:reduced}
\end{figure}

Using this representative single-basis setting, we next test the performance of the reduced protocol across system sizes. As shown in Fig.~\ref{fig:reduced}, the method maintains high classification accuracy for GHZ, W, and cluster states over $N=20,\dots,100$. The accuracy remains close to or above the 0.95 threshold for all three state families, while the number of measurement configurations grows only with the number of local windows. In particular, because $K_{\mathrm{local}}=1$, the measurement count is reduced by a factor of 81 compared with the full-basis protocol. This result confirms that a single well-chosen local basis is sufficient to classify different entanglement structures with high accuracy, and that the reduced protocol is not restricted to a specific state family.

To further test whether the reduced protocol can resolve detailed entanglement structures in large systems, we examine representative 100-qubit GHZ-type states with different block decompositions, all measured in the XZYX basis. Table~\ref{tab:100qubit} lists ten representative partitions drawn from the 30 candidate structures. These examples include both highly unbalanced partitions, where one large GHZ block is accompanied by small GHZ blocks, Bell pairs, or product qubits, and more distributed partitions containing several medium-sized GHZ blocks. The classifier achieves near-perfect or perfect accuracy for most representative structures. Even for the most difficult cases, where the partition contains several small residual components, the accuracy remains far above the random-guessing baseline. This confirms that the single-basis local joint-measurement protocol does not merely distinguish simple global classes, but can also resolve the internal block structure of large GHZ-type states.

\begin{table}[t]
  \centering
  \begin{tabular}{lc}
    \hline
    Entanglement structure & Accuracy \\
    \hline
    GHZ\_52 GHZ\_24 GHZ\_3 Bell \(\mathrm{One}_{(\times 19)}\) & 0.920 \\
    GHZ\_55 GHZ\_40 GHZ\_4 One & 0.975 \\
    GHZ\_56 GHZ\_33 GHZ\_6 GHZ\_5 & 1.000 \\
    GHZ\_56 GHZ\_43 One & 0.9111 \\
    GHZ\_64 GHZ\_22 GHZ\_12 Bell & 1.000 \\
    GHZ\_66 GHZ\_18 GHZ\_15 One & 0.9778 \\
    GHZ\_84 GHZ\_9 GHZ\_4 Bell One & 1.000 \\
    GHZ\_85 GHZ\_11 GHZ\_3 One & 1.000\\
    GHZ\_90 GHZ\_5 Bell One One One & 1.000\\
    GHZ\_97 GHZ\_3 & 1.000\\
    \hline
  \end{tabular}
  \caption{Classification accuracy for representative 100-qubit entanglement structures.}
  \label{tab:100qubit}
\end{table}

\subsection{Physical Verification and Hardware Limits}
\begin{figure}[h]
    \centering
    \includegraphics[width=0.8\linewidth]{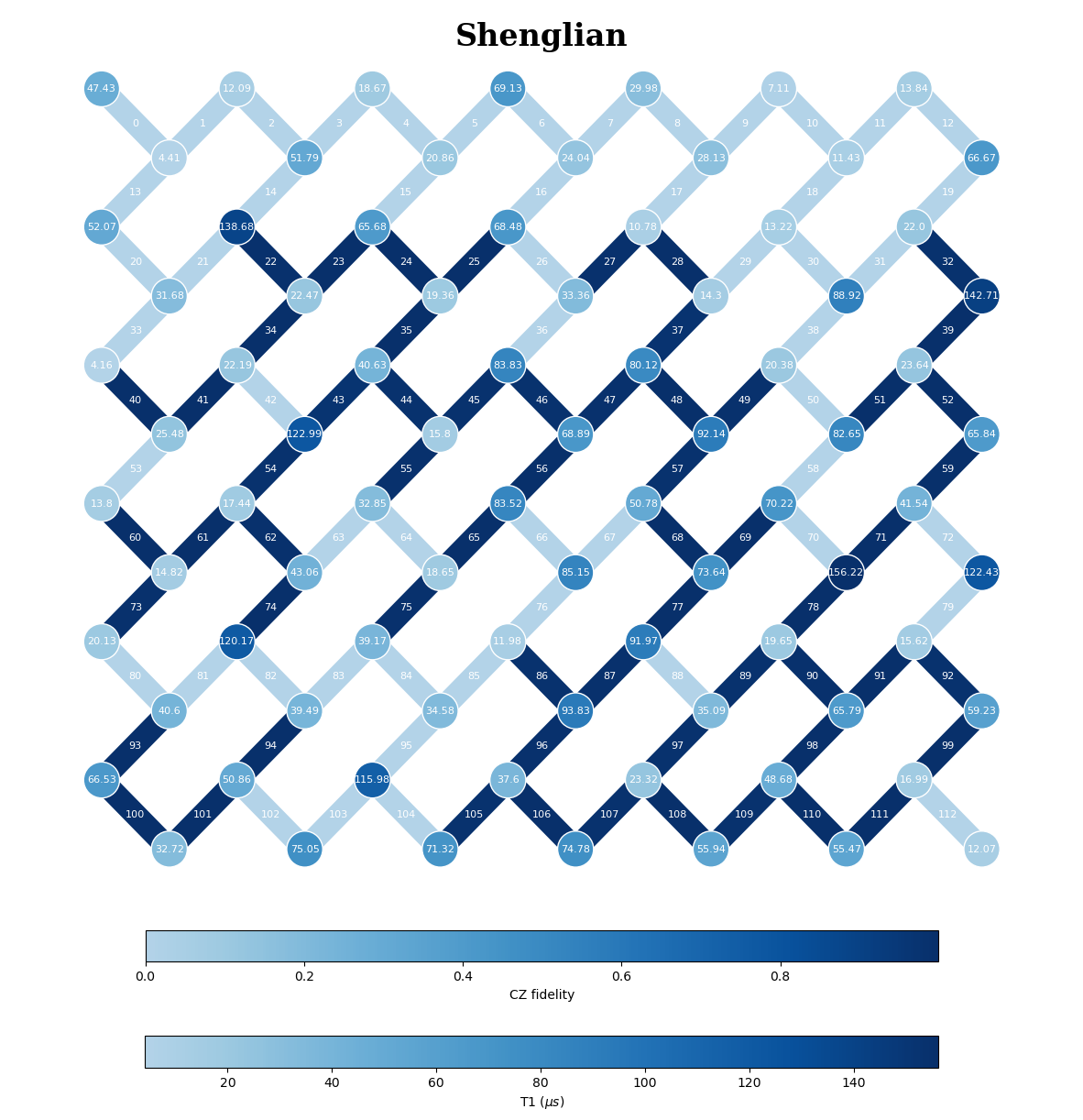}
    \caption{\textbf{Hardware topology and calibration data of the
\emph{Shenglian} superconducting quantum processor}
(Quafu cloud platform\cite{bib52}, $84$ qubits and $113$ tunable couplers). Nodes represent qubits and links represent tunable couplers; node color encodes the single-qubit relaxation time $T_1$ (in $\mu\mathrm{s}$, bottom color bar) and link color encodes the two-qubit CZ-gate fidelity (top color bar). For the calibration data used in this experiment, the median single-qubit-gate error is $1.0\times10^{-3}$ and the median two-qubit-gate error is $7.0\times10^{-3}$. }
    \label{fig:device}
\end{figure}

We now turn to hardware, where the noise is not a tunable parameter but instead grows with the depth of the state-preparation circuit. We validate the protocol on the \emph{Shenglian} superconducting quantum processor, an 84-qubit device with 113 tunable couplers, accessed through the Quafu cloud platform \cite{bib52}. Importantly, the classifier used here is the one trained on the simulation data of Sec.~2.1; no retraining on hardware data is performed. The hardware validation focuses on GHZ-type block structures, and within each window we apply the same representative local Pauli basis XZYX identified in Sec.~2.1. For the calibration used here, the median single-qubit-gate error is $1.0\times10^{-3}$ and the median two-qubit-gate error is $7.0\times10^{-3}$; the hardware topology, together with the spatially resolved two-qubit-gate fidelities and relaxation times, is shown in Fig.~\ref{fig:device}. These parameters set the constraints for circuit compilation and indicate the dominant limitation: single-qubit operations are relatively accurate, but the accumulation of two-qubit-gate error and decoherence becomes significant as the preparation circuit grows in size and depth.

\begin{figure}[t]
     \centering
    \includegraphics[width=0.95\linewidth]{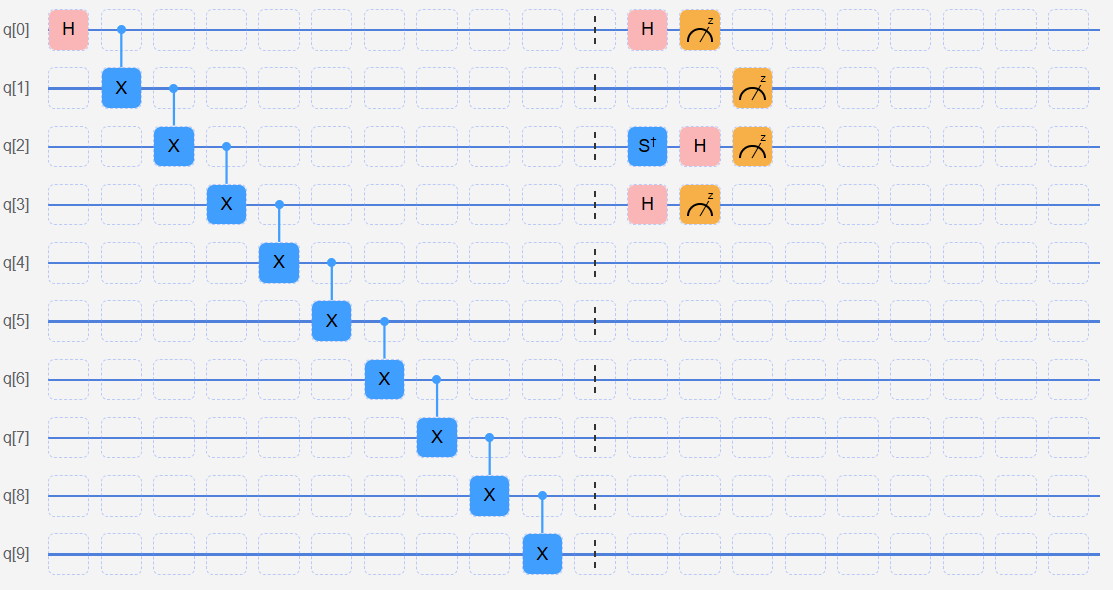}
    \caption{\textbf{Representative state-preparation circuit for a $10$-qubit GHZ state} compiled onto the native connectivity of the \emph{Shenglian} processor. All qubits are initialized in the computational basis state $\ket{0}$. A Hadamard gate first creates a coherent superposition on the initial qubit, and a sequence of entangling CNOT operations then distributes correlations across the selected qubits in a staircase pattern. After state preparation, the representative XZYX basis is applied to every window; because it satisfies $p_1=p_R=X$, all $N_{\mathrm{win}}$ windows are compatible and are read out in a single circuit. The figure illustrates the readout rotations for one window (H on the X-axis qubits, $S^{\dagger}\!H$ on the Y-axis qubit, and direct readout on the Z-axis qubit); the remaining windows reuse the identical local rotation pattern. The resulting local probability distributions are the input features for the trained classifier.}
    \label{fig:qc}
\end{figure}

For each target entanglement structure, the corresponding state-preparation
circuit is compiled onto the native connectivity of the \emph{Shenglian} processor. A representative example is the circuit used to prepare a $10$-qubit GHZ state, shown in Fig.~\ref{fig:qc}. The circuit starts from the computational basis state, applies a Hadamard gate to create a coherent superposition, and then uses a sequence of entangling operations to distribute correlations across the selected qubits. After state preparation, local measurements are performed on the relevant qubits, and the resulting local probability distributions are used as the input features for the trained classifier.

Each prediction trial on the quantum processor is implemented as follows. First, the corresponding quantum circuit is executed on the superconducting processor with $10\,000$ shots. The measurement counts from these shots are then normalized to obtain one empirical local probability distribution. This distribution is fed into the trained classifier, which outputs one predicted structural label. Therefore, one prediction trial corresponds to one independently acquired probability distribution and one classification result.

\begin{figure}[h]
    \centering
    \includegraphics[width=0.9\linewidth]{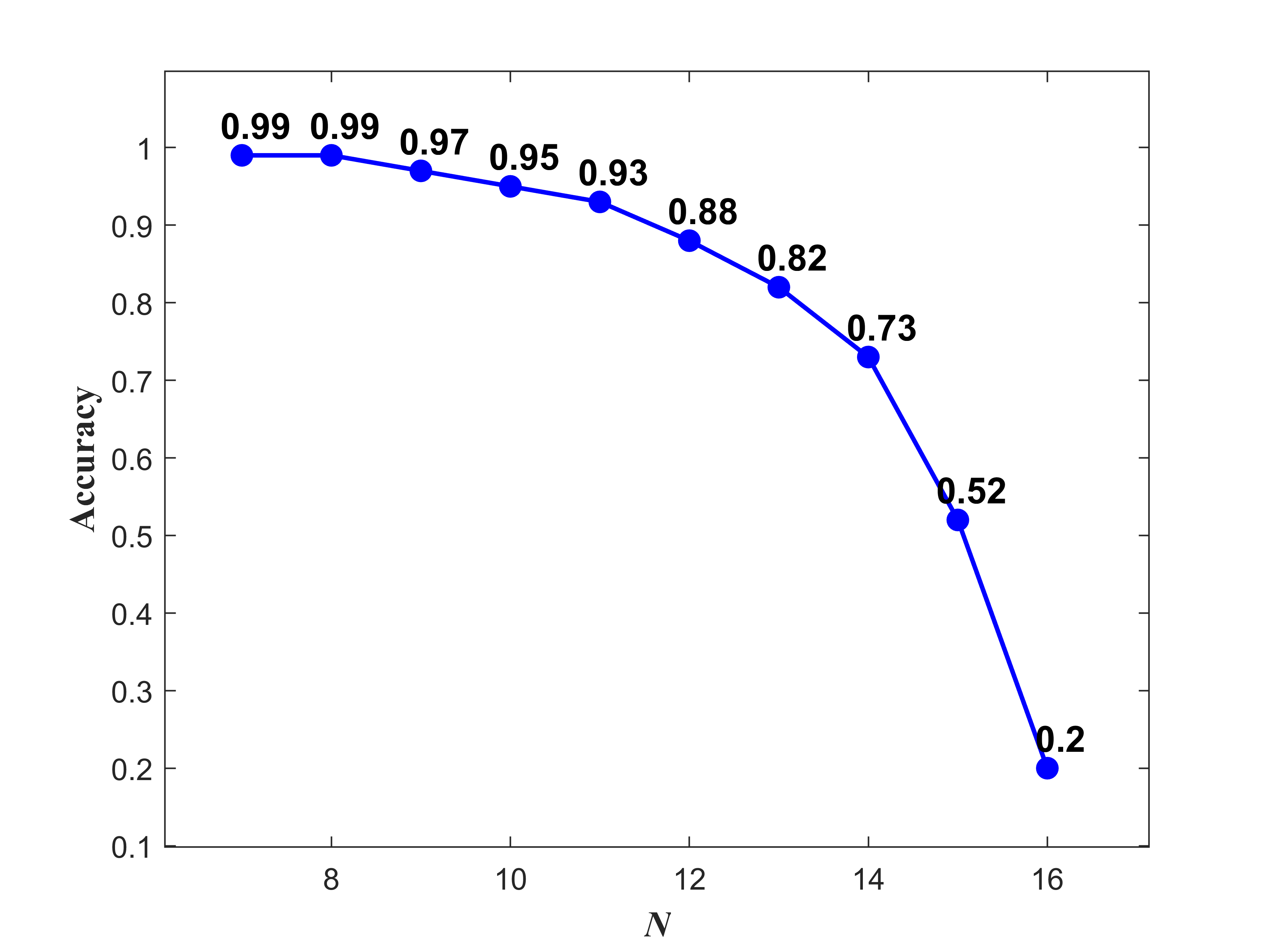}
    \caption{\textbf{Classification accuracy versus system size $N$ on the \emph{Shenglian} processor}. We test system sizes from $N=7$ to $N=16$. For each particle number, five different structural classes are considered, corresponding to different compositions of multipartite entangled blocks and single-qubit components. For each fixed particle number, $100$ prediction trials are performed in total, evenly distributed over the five structural classes ($20$ trials per class), and the accuracy is computed as the fraction of correctly predicted labels among the $100$ trials. The classification accuracy remains high for small systems ($0.99$ at $N=7$) and decreases as $N$ grows, dropping sharply for $N\gtrsim 15$ (from $0.52$ at $N=15$ to $0.2$ at $N=16$). }
    \label{fig:computer_acc}
\end{figure}

The hardware validation focuses on GHZ-type block structures. We test system sizes from $N=7$ to $N=16$, and for each particle number we consider five different structural classes, corresponding to different GHZ-block partitions combined with Bell pairs and single-qubit components .
For each fixed particle number, we perform $100$ prediction trials on the hardware in total. These $100$ trials are evenly distributed over the five structural classes, with $20$ trials for each class. The classification accuracy is computed as the fraction of correctly predicted labels among the $100$ trials. The dependence of the classification accuracy on the particle number is summarized in Fig.~\ref{fig:computer_acc}.

As shown in Fig.~\ref{fig:computer_acc}, the classifier maintains high accuracy for relatively small system sizes. The measured accuracy is $0.99$ at $N=7$, $0.99$ at $N=8$, and $0.97$ at $N=9$. As the system size increases, the accuracy decreases gradually to $0.82$ at $N=13$ and $0.73$ at $N=14$. A much stronger degradation appears at $N=15$, where the accuracy drops to $0.52$, and at $N=16$ it reaches $0.20$, which coincides with the five-class random-guess baseline of $1/5$ and thus indicates a complete loss of structural information at this size. This trend shows that the proposed local measurement features remain experimentally accessible for small and intermediate system sizes, but become increasingly fragile as the particle number, the number of two-qubit gates, and the circuit depth grow.

\begin{figure}[h]
    \centering
    \includegraphics[width=0.9\linewidth]{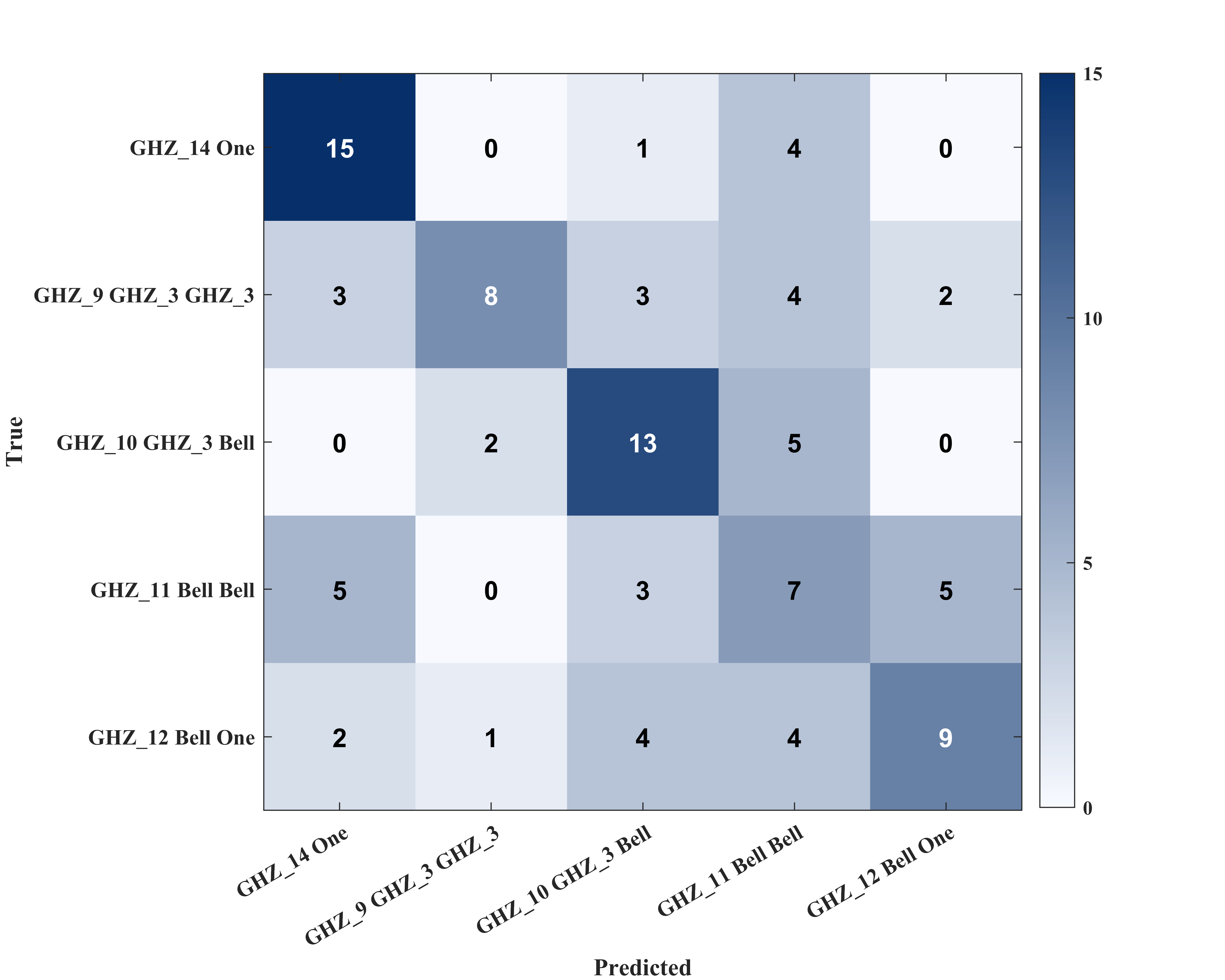}
    \caption{\textbf{Confusion matrix of the hardware results at $N=15$.} The matrix summarizes $100$ prediction trials, with each of the five structural classes contributing $20$ trials. Correct predictions lie on the diagonal, while the substantial off-diagonal weight reflects the reduced accuracy at this system size.}
    \label{fig:matrix}
\end{figure}

To analyze the degradation at $N=15$ in more detail, we examine the corresponding confusion matrix, shown in Fig.~\ref{fig:matrix}. The matrix summarizes 100 prediction trials on the hardware in total. Since the five structural classes are sampled uniformly, each true class contributes $20$ prediction trials. Thus, each row of the confusion matrix contains $20$ records, and each matrix element represents the number of trials assigned to a particular true--predicted class pair.

Ideally, most counts should concentrate on the diagonal of the confusion matrix. In the experimental result, however, visible off-diagonal entries appear for multiple true classes, indicating that a non-negligible fraction of the trials is assigned to incorrect structural labels. Although the diagonal entries remain dominant for some classes, the off-diagonal distribution shows that the local measurement features at $N=15$ are already distorted by realistic hardware noise on the quantum processor. Consequently, the classifier can no longer reliably distinguish all five entanglement structures, which is consistent with the reduced overall accuracy of $0.52$.

\begin{figure}[h]
    \centering
    \begin{subfigure}{0.3\linewidth}
        \centering
        \includegraphics[width=\linewidth]{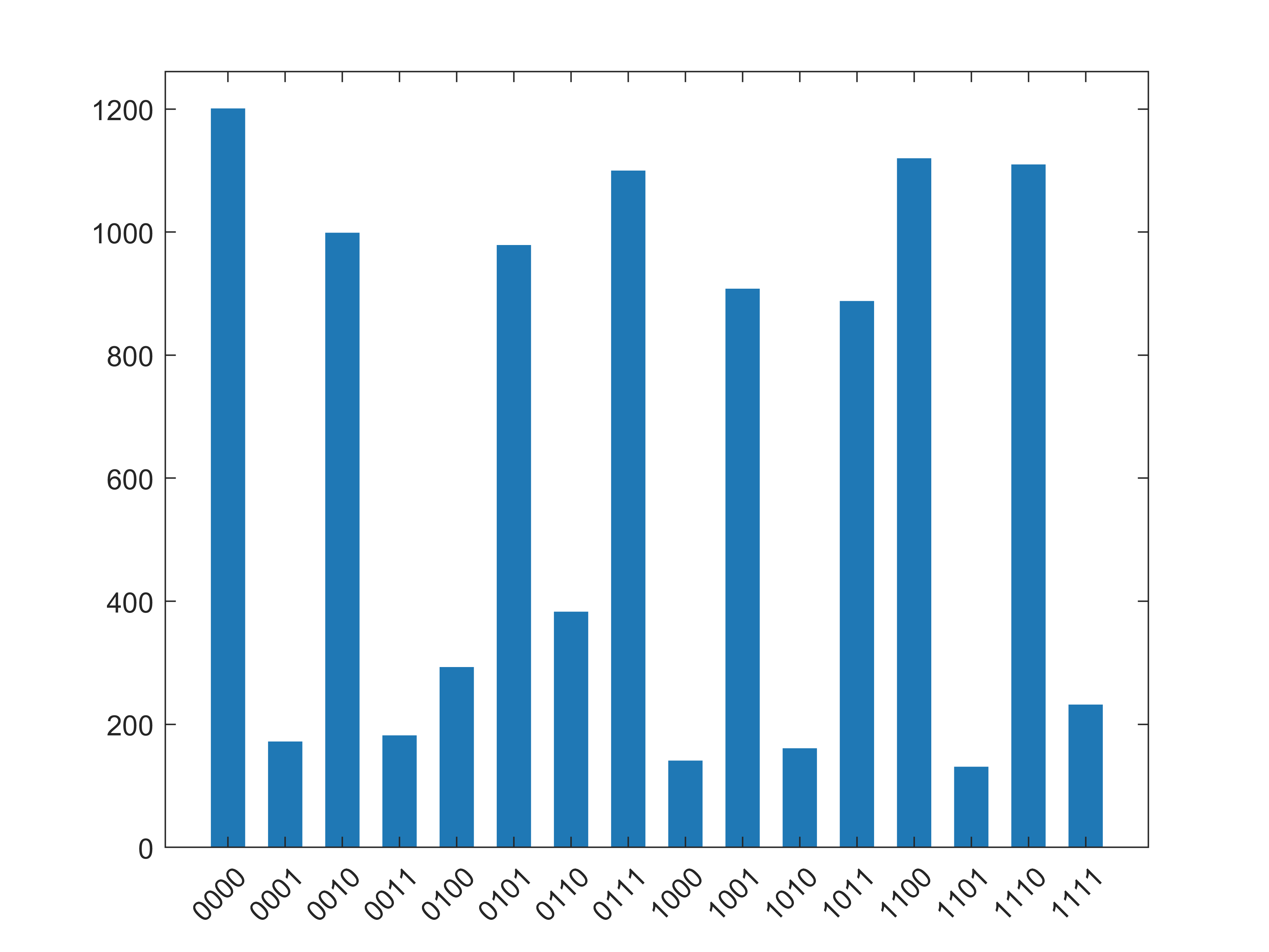}
        \caption{}
        \label{fig:result_device}
    \end{subfigure}
    \hfill
    \begin{subfigure}{0.3\linewidth}
        \centering
        \includegraphics[width=\linewidth]{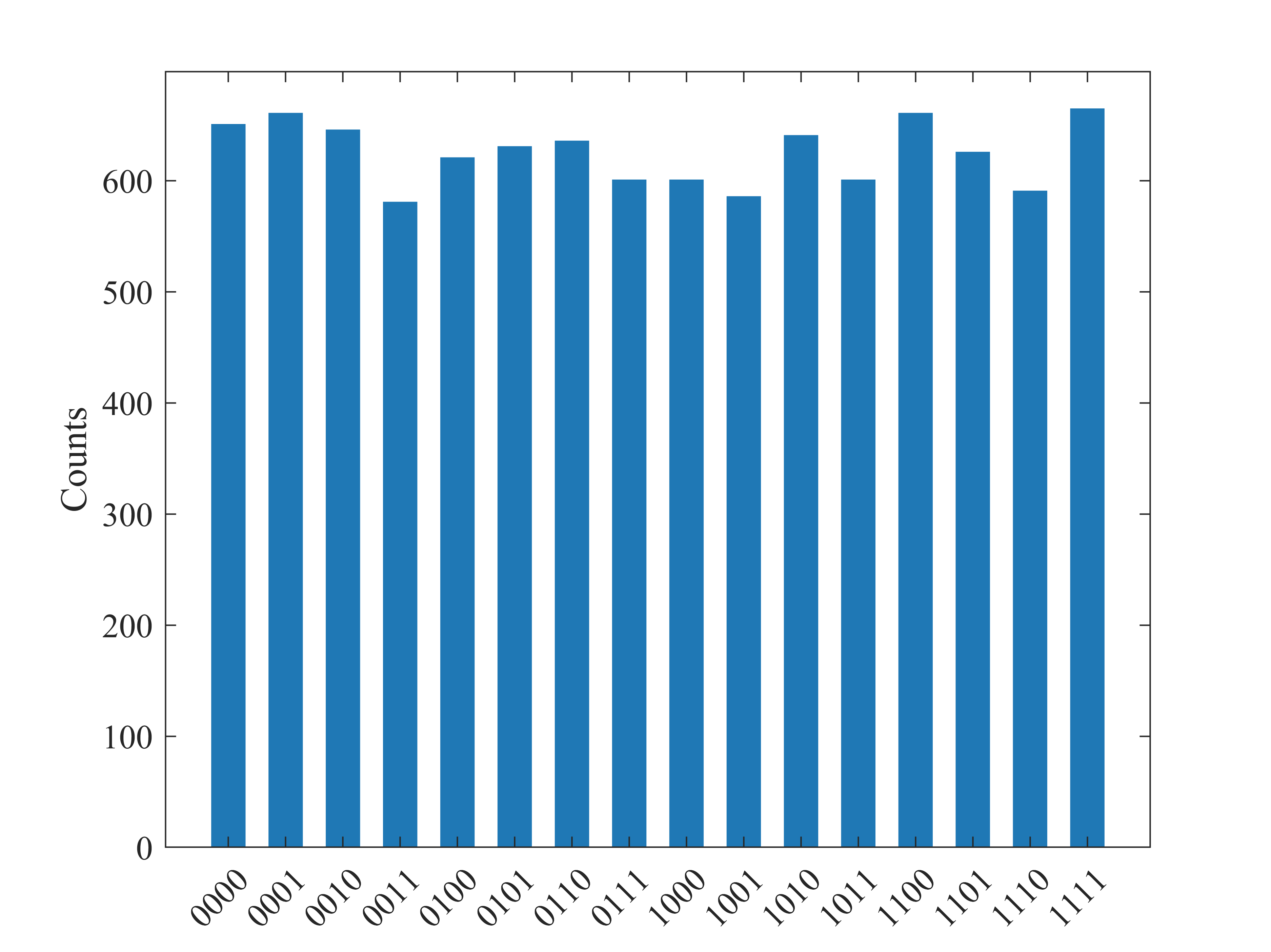}
        \caption{}
        \label{fig:result_program}
    \end{subfigure}
    \hfill
    \begin{subfigure}{0.3\linewidth}
        \centering
        \includegraphics[width=\linewidth]{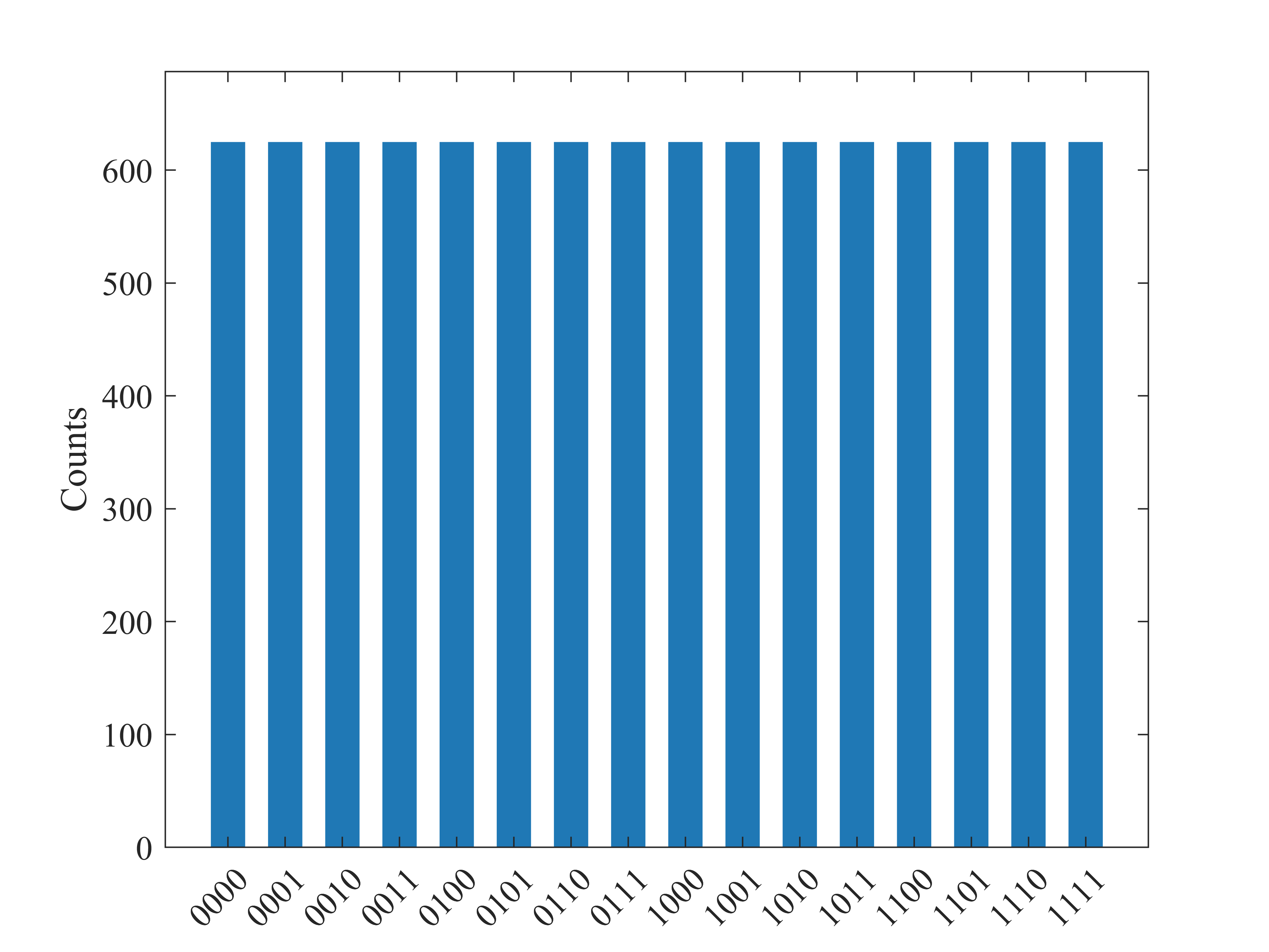}
        \caption{}
        \label{fig:result_real}
    \end{subfigure}  
    \caption{\textbf{Local measurement probability distributions under three
    conditions.} The three panels compare the distribution measured in the XZYX Pauli basis on (a) the quantum processor, (b) the program-simulated distribution, and (c) the ideal theoretical distribution. The ideal case is nearly uniform over the local bitstrings; the program simulation largely preserves this structure with only moderate deviations; the hardware result is strongly distorted, with some bitstrings over-populated and others suppressed, which weakens the distinguishability between structural classes.}
    \label{fig:result_computer}
\end{figure}

We further diagnose this behavior by comparing local measurement probability distributions obtained under three different conditions. Figure~\ref{fig:result_device} shows the distribution measured on the \emph{Shenglian} processor, Fig.~\ref{fig:result_program} shows the corresponding program-simulated distribution, and Fig.~\ref{fig:result_real} shows the ideal theoretical distribution. Since these distributions are obtained from local measurements, they should be interpreted as local probability features rather than full global-state probability distributions. In the ideal case, the local distribution displays a regular and nearly uniform profile over the measured local bitstrings. The program-simulated result largely preserves this ideal structure, with only moderate deviations arising from finite sampling or simulation details. In contrast, the measured hardware distribution is strongly distorted: some local bitstrings are over-populated, whereas others are significantly suppressed. This distributional distortion changes the local statistical features used by the classifier and weakens the distinguishability between different structural classes.

The comparison between Figs.~\ref{fig:result_device}--\ref{fig:result_real}
suggests that the dominant failure mode at $N=15$ is hardware-induced distortion of the measured local probability distributions. This fragility has a clear physical origin in the interplay between the GHZ correlations and the chosen measurement basis. Within a large GHZ block, the reduced state of an interior $R$-qubit window is exactly
\begin{equation}
\rho_{\mathrm{int}}=\tfrac{1}{2}\big(|0\rangle\!\langle 0|^{\otimes R}
+|1\rangle\!\langle 1|^{\otimes R}\big),
\label{eq:rho_int}
\end{equation}
whose XZYX outcome distribution is exactly uniform, $q(b)=2^{-R}$ for all
$b\in\{0,1\}^{R}$. Consequently the interior windows sit at a noise-insensitive fixed point: their marginals are independent of the block length and therefore carry essentially no discriminative signal about the partition. The entire classification burden thus falls on the few windows that straddle the block boundaries, which are the only ones whose marginals depend on where one coherent block ends and the next component begins.

This concentration of structural information is precisely what makes the protocol fragile under hardware noise. The boundary-spanning windows are prepared last in the entangling staircase and carry the correlations most sensitive to accumulated error, so they are exactly the features that hardware noise smears out first. Preparing large entangled states requires a sequence of two-qubit entangling operations, and given the median two-qubit-gate error of $7.0\times10^{-3}$ on the \emph{Shenglian} processor, these errors accumulate as the circuit depth increases; decoherence during circuit execution and measurement imperfections further reduce the visibility of the local correlations that distinguish different entanglement structures. Once the fragile boundary signatures are washed out, only the uninformative uniform interior marginals remain, leaving the classifier with a nearly featureless input. This explains why the experimentally measured local probability features deviate from their ideal patterns, producing the off-diagonal entries in Fig.~\ref{fig:matrix} and the abrupt accuracy drop toward the random-guess baseline observed at $N\geq 15$ in Fig.~\ref{fig:computer_acc}.

Overall, the experiments demonstrate both the feasibility and the current hardware limitations of the proposed classification protocol. For system sizes up to $N=13$, the measured local features remain sufficiently robust to support reliable classification (accuracy $\geq 0.82$ against a five-class random baseline of $0.20$); at $N=14$ the accuracy has fallen to $0.73$, marking the onset of degradation, before the sharp collapse at $N\geq 15$. For larger systems, especially around $N=15$ and beyond, the performance becomes limited by accumulated two-qubit-gate errors, decoherence, and measurement imperfections. These limitations are technical rather than conceptual: with improved two-qubit-gate fidelities, more efficient circuit compilation, readout-error mitigation, and noise-aware training, the proposed protocol is expected to extend to larger entangled systems.

\section{Methods}

\subsection{States and Data Generation}

We consider multipartite states whose entanglement structure is specified by a partition of the $N$-qubit system into mutually unentangled blocks. A structure label is denoted by
\begin{equation}
  \mathcal{P} = \{B_1,B_2,\dots,B_m\},
\end{equation}
where $B_\alpha$ is a subset of qubits with $\cup_{\alpha=1}^m B_\alpha=\{1,2,\dots,N\}$ and $B_\alpha\cap B_\beta=\varnothing$ for $\alpha\neq\beta$. The ideal target state is a tensor product of entangled states within the individual blocks,
\begin{equation}
  |\Psi_\mathcal{P}\rangle = \bigotimes_{\alpha=1}^m |\psi_{B_\alpha}\rangle .
\end{equation}
In the main benchmark each block is a GHZ state,
\begin{equation}
  |\mathrm{GHZ}_{|B_\alpha|}\rangle
  = \frac{1}{\sqrt 2}\left(|0\rangle^{\otimes|B_\alpha|}+|1\rangle^{\otimes|B_\alpha|}\right),
\end{equation}
so that different labels $\mathcal{P}$ correspond to different GHZ-block structures. This ensemble is a stringent test for local detection, because the reduced density matrices of a GHZ state carry strong classical correlations while its global phase coherence is not fully certified by local marginals alone. The task addressed here is therefore the supervised recognition of the structure label $\mathcal{P}$ within a physically motivated ensemble, rather than arbitrary entanglement certification from local data. To probe generality we also consider W-state blocks,
\begin{equation}
  |\mathrm{W}_{|B_\alpha|}\rangle
  = \frac{1}{\sqrt{|B_\alpha|}}\sum_{j\in B_\alpha}|0\cdots 1_j\cdots 0\rangle,
\end{equation}
and cluster-state (graph-state) blocks on the corresponding subsets of qubits. For each system size $N$, structure labels are generated from random integer partitions of $N$, and the blocks are instantiated as GHZ, W, or cluster components combined with Bell pairs and single-qubit (product) components. Because the labels are sampled independently at each $N$, the difficulty of the resulting ensemble varies somewhat across system sizes; this is the origin of the non-monotonic accuracy fluctuations reported in the main text.

To emulate realistic state-preparation imperfections while keeping the ideal block structure intact, all training and test data in the main experiments are generated with a rotation-gate perturbation. After each block is instantiated in its ideal target state, every qubit is acted on by an independent random local unitary
\begin{equation}
  U_i = R_z(\theta_z)\,R_y(\theta_y)\,R_x(\theta_x),
\end{equation}
with the three angles drawn uniformly from $(0,\delta)$ and $\delta = 15\pi/100$. For Bell pairs the perturbation is applied directly to the two-qubit state vector; for large blocks it is applied site by site to the matrix-product-state tensors, leaving the bond dimension unchanged. 

For the stress test of Fig.~\ref{fig:stress}, we additionally apply two physically distinct and independently controlled channels defined \emph{at the
level of the quantum state}: a preparation-stage depolarizing channel and a global white-noise channel. Both channels are specified physically on the state
[Eqs.~\eqref{eq:depol} and \eqref{eq:white}]; however, because each admits an
\emph{exact} closed-form action on the $R$-qubit window marginals, we evaluate
them directly on the clean measurement statistics without ever reconstructing
the reduced density matrix or the global state $\rho$.

During state preparation, every qubit is subjected to an independent single qubit depolarizing channel with Kraus operators $\{\sqrt{1-p_d}\,I,\ \sqrt{p_d/3}\,X,\ \sqrt{p_d/3}\,Y,\ \sqrt{p_d/3}\,Z\}$, i.e.
\begin{equation}
\mathcal{E}_d(\rho) = (1-p_d)\,\rho
      + \frac{p_d}{3}\left(X\rho X + Y\rho Y + Z\rho Z\right),
\label{eq:depol}
\end{equation}
where $p_d$ is the total single-qubit error probability. Since $\mathcal{E}_d$
is unital and self-dual, its dual action on a computational-basis effect is $\mathcal{E}_d^{\dagger}(\lvert s\rangle\!\langle s\rvert)
 = (1-\tfrac{4p_d}{3})\,\lvert s\rangle\!\langle s\rvert
 + \tfrac{2p_d}{3}\,I$. Two consequences follow. First, on any qubit \emph{outside} the window $\mathcal{E}_d^{\dagger}(I)=I$, so applying the channel to traced-out qubits leaves the window marginal unchanged; the block wise (small blocks, $n_b\le 2$) and site-wise (large blocks, on the matrix-product operator) description of the channel therefore acts trivially there. Second, on each qubit \emph{inside} the window it reduces to the single-axis stochastic map
\begin{equation}
P'(s_i) = \Big(1-\tfrac{4p_d}{3}\Big)P(s_i)
        + \tfrac{2p_d}{3}\big[P(s_i)+P(\bar{s}_i)\big],
\label{eq:depol_marginal}
\end{equation}
where $\bar{s}_i = 1-s_i$ denotes the bit complement, applied independently along the $R$ axes of the window probability tensor. Thus the marginal-level evaluation reproduces the state-level channel of Eq.~\eqref{eq:depol} exactly.

The white-noise component is a genuine global channel that mixes the $N$-qubit state with the maximally mixed state,
\begin{equation}
\rho \longrightarrow \rho' = (1-p_w)\,\rho + p_w\,\frac{I}{2^{\,N}}.
\label{eq:white}
\end{equation}
Although this is defined globally, its action on any local $R$-qubit window is exact. For a window $W$ with $\lvert W\rvert = R$, the POVM element associated
with outcome $s$ is $M_s = U^{\dagger}\lvert s\rangle\!\langle s\rvert U \otimes I_{\mathrm{rest}}$, and

\begin{equation}
P'(s) = \mathrm{Tr}[\rho' M_s]
      = (1-p_w)\,\mathrm{Tr}[\rho M_s]
      + p_w\,\mathrm{Tr}\!\Big[\tfrac{I}{2^{\,n}} M_s\Big]
      = (1-p_w)\,P(s) + p_w\,\frac{1}{2^{\,R}},
\label{eq:white_marginal}
\end{equation}
where we used
$\mathrm{Tr}[(I/2^{\,N})M_s] = (1/2^{\,N})\cdot 1\cdot 2^{\,N-R} = 1/2^{\,R}$,
since the reduction of the maximally mixed state to any $R$-qubit window is
$I/2^{\,R}$. This is an exact identity, not an approximation: on every local window, the global white-noise channel is equivalent to mixing the clean marginal distribution with the uniform distribution over the $2^{\,R}$ outcomes. The mixture is automatically normalized, $(1-p_w)\sum_s P(s) + p_w\cdot 2^R\cdot 2^{-R} = 1$. In the stress test the two strengths are swept independently, $p_d\in[0,0.2]$ and $p_w\in[0,0.4]$, and each configuration is evaluated separately.

To generate data at large $N$ we compute the local distributions by tensor-network contraction rather than explicit density-matrix reconstruction. The block-structured states admit efficient matrix-product-state (MPS) representations: GHZ blocks have an exact MPS of small bond dimension when the qubits in a block are contiguous, and W and one-dimensional cluster blocks likewise admit compact MPS descriptions of bond dimension $\chi=2$. The reduced density matrix $\rho_{W_j}$ of a window is obtained by contracting the environment tensors outside the window into boundary tensors and evaluating the resulting $R$-site effective network, without ever forming the full $2^N\times 2^N$ density matrix. For bond dimension $\chi$, the cost of a local $R$-qubit marginal scales polynomially in $\chi$ and exponentially only in the fixed window size $R$; since $R$ is held constant, the cost of generating all local features grows linearly with $N_{\mathrm{win}}$. This tensor-network realization is thus aligned with the measurement scaling of the protocol.

\subsection{Local Measurement and Basis Selection}

The central ingredient of the protocol is an overlapping local representation of multipartite correlations. For an $N$-qubit chain we choose a window size $R$ and a step $s\le R$ so that neighbouring windows overlap or touch and the chain is fully covered. The first $N_{\mathrm{win}}-1$ windows are placed at equal spacing,
\begin{equation}
  W_j = \{1+(j-1)s,\dots,1+(j-1)s+R-1\}, \quad j=1,\dots,N_{\mathrm{win}}-1,
\end{equation}
and the final window is anchored at the right boundary, $W_{N_{\mathrm{win}}}=\{N-R+1,\dots,N\}$, so that every qubit is covered. The number of windows is
\begin{equation}
  N_{\mathrm{win}} = \left\lceil\frac{N-R}{s}\right\rceil + 1 = O(N),
\end{equation}
for fixed $R$ and $s$. Throughout this work we use $R=4$ and $s=3$, which were found to give reliable classification at low measurement cost across all tested system sizes and state families; the method is not sensitive to the precise choice within a moderate-overlap regime.

Within each window $W_j$ we measure a local Pauli basis specified by a string
$\mathbf{p}=(p_1,\dots,p_R)$ with $p_\ell\in\{X,Y,Z\}$, obtaining the outcome distribution
\begin{equation}
  \mathbf{q}_{j,\mathbf{p}} = \big(q_{j,\mathbf{p}}(\mathbf{b})\big)_{\mathbf{b}\in\{0,1\}^R},
  \quad q_{j,\mathbf{p}}(\mathbf{b}) = \mathrm{Tr}[\rho_{W_j}\,\Pi_\mathbf{p}(\mathbf{b})],
\end{equation}
where $\rho_{W_j}$ is the reduced density matrix on $W_j$ and
$\Pi_\mathbf{p}(\mathbf{b})=\bigotimes_{\ell=1}^R|\phi^{b_\ell}_{p_\ell}\rangle\langle\phi^{b_\ell}_{p_\ell}|$
is the product projector of basis $\mathbf{p}$ and outcome $\mathbf{b}$. The dimension of this local distribution is $2^R$, independent of $N$. The complete local representation is the concatenation over all selected bases and windows,
\begin{equation}
  \mathbf{x} = \bigoplus_{j=1}^{N_{\mathrm{win}}}\bigoplus_{\mathbf{p}\in\mathcal{S}_K}
  \mathbf{q}_{j,\mathbf{p}},
\end{equation}
where $\mathcal{S}_K$ is the selected set of local bases with $K_{\mathrm{local}}=|\mathcal{S}_K|$.
The same basis set is applied to every window, which keeps the measurement design translationally uniform.

The full set of local Pauli bases in an $R$-qubit window contains $3^R$ elements; measuring all of them is unnecessary when only the structure label is required. As shown in the main text, the structural information is redundantly distributed across bases, so that already a single local basis ($K_{\mathrm{local}}=1$) suffices in most cases; the per-basis sweep of Fig.~\ref{fig:perbasis} confirms that the large majority of individual bases are informative and only a small minority carry little structural signal. Guided by the stress test of Fig.~\ref{fig:stress}, we fix the representative single basis per state family. To guarantee that overlapping windows agree on their shared qubits, so that the entire chain can indeed be read out in a single measurement configuration, we additionally require the boundary-matching condition $p_1=p_R$. Among the high-accuracy candidates this yields XZYX for GHZ ($p_1=p_R=X$). For W and cluster states we adopt the highest-accuracy bases that also satisfy $p_1=p_R$, namely YXYX and ZZZX; the previously reported YXYX and ZZZX violate $p_1=p_R$ and would require two interleaved configurations, which we no longer use. Because the choice does not depend on $N$, a basis fixed at small $R$ can be reused for larger systems.

Three quantities must be distinguished. The number of distinct measurement configurations, the physically independent circuits run on hardware, is determined by $\mathcal{S}_K$ and is independent of $N$, since the same local basis set is reused in every window; for $K_{\mathrm{local}}=1$ it is a single configuration for the entire chain. The outcome space of each window is fixed at $2^R$, in contrast to the $2^N$ outcomes of a global $N$-qubit Pauli measurement. The feature dimension
\begin{equation}
  D = N_{\mathrm{win}}\,K_{\mathrm{local}}\,2^R
\end{equation}
grows only linearly with $N$. Thus both the experimental sampling cost and the classifier input scale with the fixed window size rather than the full Hilbert-space dimension.

\subsection{Classifier and Training}

The concatenated local representation $\mathbf{x}$ is the input to a neural-network classifierthat outputs the predicted structure label $\mathcal{P}$. The classifier is a compact fully connected network that takes the length-$D$ local-feature vector [Eq.~(17)] as input, followed by two hidden layers of width 256 and 128, each with batch normalization, a ReLU activation, and (for the first layer) dropout with rate 0.3; a final linear layer maps the resulting representation to the $N_{\mathrm{class}}$ structure labels. The network is trained by minimizing the cross-entropy loss with the Adam optimizer for up to 300 epochs, using early stopping (patience 30) and learning-rate scheduling on a held-out validation set obtained from an 80/20 split. Full architectural and training details, together with the datasets and code, are provided in the Supplementary Information and the associated repository. We cast entanglement-structure detection as a supervised multi-class classification problem, in which each class corresponds to one admissible block partition $\mathcal{P}=\{B_1,\dots,B_m\}$ of the $N$-qubit chain. The candidate partitions are drawn as a fixed set of $N_{\mathrm{class}}$ distinct structures, and the number of classes fixes the corresponding random-guess baseline $1/N_{\mathrm{class}}$ against which all reported accuracies must be read. Two settings are used throughout this work. (i) In the large-scale numerical benchmarks ($N=20,30,\dots,100$), the classifier discriminates among $N_{\mathrm{class}}=30$ candidate block partitions, so that the random-guess baseline is $1/30\approx 3.3\%$; the simulated GHZ-, W- and cluster-block accuracies, as well as the 100-qubit results in Table~\ref{tab:100qubit}, refer to this 30-class task. (ii) In the superconducting-hardware experiments ($N=7,\dots,16$), where the state-preparation cost restricts the number of structures that can be reliably compiled, each particle number is assigned $N_{\mathrm{class}}=5$ structural classes, giving a random-guess baseline of $1/5=20\%$. For each $N$ the 100 prediction trials are distributed uniformly over these five classes (20 trials per class), and the accuracy is the fraction of correctly predicted labels.

Detection performance is quantified by the classification accuracy on a held-out test set of independently generated states from the same ensemble. In the hardware experiments, the selected local Pauli bases are implemented by single-qubit basis rotations followed by computational-basis readout; the empirical local distributions are estimated from finite-shot counts and processed in the same window order as the numerical representation. No full state reconstruction is performed at any stage.

\section{Discussion}\label{sec:discussion}

We have shown that the entanglement structure of a large multipartite state can be recognized from local joint measurements, without reconstructing the full state or performing global Pauli measurements. The method rests on a single physical fact: for structured multipartite states, the block partition is redundantly encoded in local correlations. This redundancy has two practical consequences that run through all of our results. First, a single local basis per window ($K_{\mathrm{local}}=1$) already achieves high accuracy, so the number of distinct measurement configurations is independent of system size and the feature dimension grows only linearly with $N$. Second, because the structural information is shared across many local bases rather than concentrated in a few, and because almost any reasonable local basis works, the method does not depend on a finely tuned measurement choice, making it robust and straightforward to deploy on hardware.

It is important to be precise about the nature of the task and its advantage. For idealized pure states, block boundaries appear as sharp jumps in short-range correlators, and the recognition problem is comparatively easy; a naive correlation-based rule can already locate the partition. The regime that matters for quantum computers is different. Under realistic noise the sharp signatures are smeared. The learning-based classifier exploits the full local outcome distribution across redundant windows and remains accurate throughout the nominal operating regime (Figs.~5 and 7). We expect this to be more robust than simple correlation-threshold heuristics, though a systematic head-to-head comparison is left to future work.

Our approach is also distinct in \emph{kind} from randomized-measurement techniques such as classical shadows. Those methods estimate expectation values of observables, i.e.\ linear functionals of the state. For such linear functionals, size-independent sampling guarantees are known.Entanglement structure detection instead asks which separability class a state belongs to, a membership question over structured convex sets, and the shadow guarantees do not transfer directly. Relative to our earlier global-measurement approach, which records $2^N$-dimensional outcome distributions and was limited to $N\lesssim 19$, the present local formulation caps the outcome space at $2^R$, keeps the number of local windows independent of $N$, and thereby extends structure detection to systems of up to $100$ qubits in simulation.

These advantages come with a well-defined operational limit. Because our candidate ensembles are constructed to be locally distinguishable, which is consistent with the perfect accuracy obtained from the full basis set, the failures observed on hardware are technical rather than representational in origin. For larger prepared states, accumulated two-qubit-gate error and increased circuit depth distort the measured distributions and wash out the correlation features, bounding the system size that can currently be reliably validated on hardware to $13$ qubits (with the onset of degradation at $N = 14$ and a sharp collapse at $N \geq 15$). This limitation is expected to ease with improved gate fidelity, readout-error mitigation, and noise-aware training, which are natural directions for extending the demonstrated hardware reach.

In summary, local correlation signatures provide a robust, experimentally accessible, and scalable representation of multipartite entanglement structure. Using a size-independent number of local windows and a feature dimension that grows only linearly with $N$, the method classifies GHZ-, W-, and cluster-type structures among many candidate partitions with high accuracy up to $100$ qubits in noisy simulation, and it is validated on a superconducting processor for systems up to $13$ qubits. By trading the goal of full state characterization for the more targeted problem of structure recognition, and by exploiting the redundancy of local correlations, this work offers a practical route to entanglement-structure characterization on near term quantum devices.

\backmatter


\bmhead{Acknowledgements}

This work was supported by the National Natural Science Foundation of
China (Grant No. 12504433) and the Beijing Information Science and Technology
University Free Exploration Project (bistu71E2510906).

\section*{Data and Code Availability}
The simulated datasets, the trained classifier weights, and the code used to
generate the local-measurement features and to reproduce all figures are
available at [https://github.com/liray123/Large-Scale-Entanglement-Structure-Detection-in-100-Qubit-Systems-via-Local-Joint-Measurements.git]. The raw hardware measurement counts from the
\emph{Shenglian} processor are available from the corresponding author upon
reasonable request.







\bibliography{sn-bibliography}

\end{document}